\documentclass[aps,prd,letterpaper,reprint,superscriptaddress,nofootinbib,preprintnumbers,longbibliography,floatfix]{revtex4-1}

\usepackage{graphicx}   
\usepackage{color}      
\usepackage[dvipsnames]{xcolor}      
\usepackage{hyperref}   

\usepackage{textcomp}
\usepackage{tikz}
\usepackage{ifthen}
\usepackage[utf8]{inputenc} 
\usepackage{bm}
\usepackage{multirow}
\usepackage{amsmath}
\usepackage{amssymb}

\usepackage[caption=false, labelformat=simple]{subfig}
\newcommand{\subfigref}[1]{\protect\subref{#1}}

\DeclareMathOperator{\E}{\mathrm{E}}
\DeclareMathOperator{\logit}{\mathrm{logit}}

\newcommand{\jet}{\mathbf{J}}
\newcommand{\trim}{\mathrm{trim}}
\newcommand{\figref}[1]{Fig.~\ref{#1}}
\newcommand{\Figref}[1]{Figure~\ref{#1}}
\renewcommand{\eqref}[1]{Eq.~\ref{#1}}

\newcommand{\tableref}[1]{table~\ref{#1}}
\newcommand{\Tableref}[1]{Table~\ref{#1}}
\newcommand{\sectionref}[1]{Sec.~\ref{#1}}
\newcommand{\Sectionref}[1]{Section~\ref{#1}}

\newcommand{\xmorph}{x_{\mathrm{morph}}}
\newcommand{\xtrim}{x_{\trim}}
\newcommand{\xleadingjet}{x_{\jet_1}}
\newcommand{\xkin}{x_{\mathrm{kin}}}
\newcommand{\ximage}{x_{\mathrm{image}}}

\newcommand{\pythia}{\texttt{PYTHIA8}}
\newcommand{\herwig}{\texttt{HERWIG7}}

\renewcommand{\vec}[1]{\mathbf{#1}}

\newcommand{\MF}{\mathrm{MF}}

\begin{document}
\title{
Morphology for Jet Classification
}

\preprint{KEK-TH-2266}

\author{Sung Hak Lim}
\email{sunghak.lim@rutgers.edu}
\affiliation{NHETC, Department of Physics and Astronomy,
Rutgers University, Piscataway, NJ 08854, USA}

\author{Mihoko M. Nojiri}
\email{nojiri@post.kek.jp}
\affiliation{Theory Center, IPNS, KEK,
1-1 Oho, Tsukuba, Ibaraki 305-0801, Japan}
\affiliation{The Graduate University of Advanced Studies (Sokendai),
1-1 Oho, Tsukuba, Ibaraki 305-0801, Japan}
\affiliation{Kavli IPMU (WPI), University of Tokyo,
5-1-5 Kashiwanoha, Kashiwa, Chiba 277-8583, Japan}

\begin{abstract}
We introduce a jet tagger based on a neural network analyzing the Minkowski Functionals (MFs) of pixellated jet images. 
The MFs are geometric measures of binary images, and they can be regarded as a generalization of the particle multiplicity, which is an important quantity in jet tagging.
Their changes by dilation encode the jet constituents' geometric structures that appear at various angular scales. 
We explicitly show that this analysis using the MFs together with mathematical morphology can be considered a constrained convolutional neural network (CNN). 
Conversely, CNN could model the MFs in a certain limit, and we show their correlation in the example of tagging semi-visible jets emerging from the strong interaction of a hidden valley scenario. 
The MFs are independent of the IRC-safe observables commonly used in jet physics. 
We combine this morphological analysis with an IRC-safe relation network which models two-point energy correlations. 
While the resulting network uses constrained input parameters, it shows comparable dark jet and top jet tagging performances to the CNN. 
The architecture has significant computational advantages when the available data is limited.  
We show that its tagging performance is much better than that of the CNN with a small number of training samples. 
We also qualitatively discuss their parton-shower model dependency. 
The results suggest that the MFs can be an efficient parameterization of the IRC-unsafe feature space of jets.
\end{abstract}

\maketitle

\tableofcontents

\section{Introduction}

The large hadron collider (LHC) has provided significant opportunities for searches of new physics beyond the standard model.
In the future extensions of the LHC, the sign of new physics may appear behind high $p_T$ jets originating from the massive gauge bosons, top quarks, or Higgs bosons.
Those boosted jets can be identified by examining jet substructures \cite{Butterworth:2008iy}, and recently, there are considerable efforts on using deep learning for tagging them \cite{Almeida:2015jua,deOliveira:2015xxd,Louppe:2017ipp,Cheng:2017rdo,Qu:2019gqs}.

The jet classification relies on substructures of jets from boosted massive particles.
\cite{Butterworth:2008iy,Dasgupta:2013ihk,Larkoski:2014wba, Krohn:2009th,Ellis:2009su,Ellis:2009me,Dreyer:2018tjj}.
The quantification of those features may be performed with jet shape variables, such as $n$-subjettiness \cite{Thaler:2010tr}, or  energy correlation functions \cite{Larkoski:2013eya}.
In particular, these variables are often described by a set of $n$-point energy correlators \cite{Tkachov:1995kk,Komiske:2017aww}, which is a basis of jet substructure variables with infrared and collinear (IRC) safety conditions.

On the other hand, counting variables, such as the number of charged tracks \cite{Gallicchio:2011xq}, are yet another type of discriminative variable in jet tagging, but there is some subtlety in predicting them by QCD because they are not IRC safe.
Those IRC unsafe features are often empirically modeled in event simulations.  
The predicted distribution often has a sizable deviation from the experimental data.
We have to use them carefully, so that classification models are not biased to a particular simulators.

Meanwhile, these feature engineering may be replaced with deep learning. 
For example, convolution-based networks \cite{deOliveira:2015xxd,Qu:2019gqs} using (pixelated) particle distributions, and recurrent neural networks \cite{Louppe:2017ipp,Cheng:2017rdo} using predefined sequence of particles are known for good jet tagging performance \cite{Kasieczka:2019dbj}.

Those networks can represent a wide variety of functions, and they cover the high-dimensional phase space of inputs. 
However, some phase space of the training sample may be underrepresented by a finite number of samples, and the jet taggers based on them require high-quality samples to get the best performance.
Because of that, it is often necessary to use dimensionality reductions, such as introducing bottlenecks in the middle of their architecture.
But those reduction techniques may not respect the physical constraints of the system, and explaining the outputs in domain-specific languages is less straightforward.
Intensive post-analysis is often required in order to get an insight from the trained networks.

In this regard, starting from physics-inspired inputs and network architectures \cite{Komiske:2018cqr,Chakraborty:2019imr,Chakraborty:2020yfc,Andreassen:2018apy,Andreassen:2019txo} has advantages over the general functional model trained on primitive inputs in controllability and interpretability.
For example, the energy flow network (EFN) \cite{Komiske:2018cqr,NIPS2017_f22e4747} and the relation network (RN) \cite{Chakraborty:2019imr,Chakraborty:2020yfc, raposo2017discovering,NIPS2017_7082} is known for its good tagging performance under the IRC-safe constraints \cite{Kasieczka:2019dbj,Chakraborty:2020yfc}. 
If those constrained models cover all the relevant features for solving the given problem, the model will have equal performance compared to the general-purpose models \cite{Chakraborty:2019imr,Chakraborty:2020yfc}.
So far, the networks covering IRC safe variables are well studied, 
but constrained models for IRC unsafe variables are not available yet. 
We need architectures bridging between general models and IRC unsafe variables.

Although deep learning models that systematically cover those IRC unsafe variables are not available, there are several frameworks based on multiplicities in coarse-graining
\cite{Davighi:2017hok}, 
dilation and Minkowski functionals \cite{Chakraborty:2020yfc}, 
and Delaunay triangulation and its topology \cite{Li:2020jdb}.
In this paper, we thoroughly reintroduce the approach in \cite{Chakraborty:2020yfc} in terms of the mathematical morphology and integral geometry, build a constrained model for the IRC unsafe variables, and show its analytic representation in the large network width limit.
 
This paper is organized as follows. 
In \sectionref{sec:mf}, we introduce the morphological analysis on jet images using Minkowski functionals (MF), which is a generalization of counting variables by using its abstract algebraic features.
We point out that the MFs can be represented bya chain of convolutions of the jet images and $2\times 2$ filters, and therefore, convolutional neural networks (CNN) can utilize it.

\Sectionref{sec:irc_safe_net} reviews the two IRC-safe energy correlator-based networks
which may provide complementary information to the MFs.
In the case of jet image analysis, we show that the RN simplifies to a multilayer perceptron (MLP) taking a two-point energy correlation $S_2(R)$, which is 
{an energy-weighted count of pairs of jet constituents} at a given angular scale. 
On the other hand, the EFN is an MLP taking the jet image itself, where the jet image is an energy flow with a finite angular resolution.

In \sectionref{sec:combined_network}, we introduce a modular architecture combining the morphological analysis and RN (or EFN).
We simply combine outputs of each network using another MLP to get the final outputs.
We are going to compare the RN (or EFN) augmentedwith the morphological analysis, against the baseline CNN.

\Sectionref{sec:benchmark_tagging} is devoted to the jet tagging performance between the combined setup using RN or EFN and the CNN.
We consider two benchmark scenarios: tagging semi-visible jets \cite{Cohen:2015toa}, and top jets.
By using the semi-visible jet tagging example, we show that CNN can learn the distinctive feature of the MFs when the difference in the MF distributions between the signal and background is significant. 
Besides, our combined architectures and CNN augmented by the MFs show better performance than baseline CNN. 
This contradicts the observation that CNN can represent the MFs.
These performance differences may be originated from the finite network size effects and regularization.

\Sectionref{sec:training_performance} discusses computational advantages of our constrained architecture compared to those of CNN.
We show that the constrained architecture has better generalization performance when the number of training samples are small.
We also point out that our setup is faster and memory-efficient because of lower computational complexity.
In short, the MFs can efficiently represent IRC-unsafe information about the jet constituents.

Existing event simulation tools such as \texttt{Pythia} \cite{Sjostrand:2014zea} and \texttt{Herwig} \cite{Bellm:2015jjp,Bahr:2008pv} predict different soft particle distributions. 
Therefore, special care is needed to estimate the classification performance using simulated datasets.
\Sectionref{sec:ps_and_mf} shows the generator dependence of jet constituent distributions in terms of MFs and describes the connection of qualitative features to the shower algorithms.

\section{Generalization of Counting Variables in Jet Physics}
\label{sec:mf}

In order to generalize the counting variables, such as particle multiplicities, we need to 
introduce the mathematical concepts called \emph{valuation}. 
The particle multiplicities, which is essentially the number of elements in a set, has the following characteristic property for union and intersection of two sets of particles, $A$ and $B$,
\begin{equation}
    n(A \cup B) = n(A) + n(B) - n(A \cap B).
\end{equation}
This abstract mathematical feature is called valuation in measure theory.
For example, area of a region is a valuation.
It would be worth exploring the space of valuations to generalize the counting variables,
and Minkowski functionals and Hadwiger's theorem are the important tools for that.

\subsection{Minkowski Functionals and Hadwiger's theorem}

The MFs of the jet constituents are the key characteristics for analyzing the space of valuations of jet substructures.
Since we are going to analyze jet images on the pseudorapidity-polar coordinate plane, we will focus on discussing the MFs for two-dimensional Euclidean space $\mathbb{R}^2$.
We also denote the coordinate vector as $\vec{R}=(\eta,\phi)$.

For a closed and bounded set $S$ in  $\mathbb{R}^2$, there are the three MFs: area $A$, boundary length $L$, and Euler characteristic $\chi$.
They can be expressed as the integral of local features of $S$ as follows,
\begin{equation}
\label{eqn:mfs:continuum}
A 
=
\int_S d^2 \vec{R}, 
\quad
L
= 
\int_{\partial S} d \vec{R},
\quad 
\chi
=
\frac{1}{2\pi} \int_{\partial S} \kappa \, d \vec{R} 
\end{equation}
where $\kappa$ is the curvature of the boundary $\partial S$.
The integral representation of the Euler characteristic is the Gauss-Bonnet theorem.

The MFs are useful measures because of its completeness.
Hadwiger's theorem \cite{hadwigeb1956integralsatze, https://doi.org/10.1112/S0025579300014625} states that these three functionals are complete basis for the translation and rotation invariant valuations of convex bodies
where the convex body is a closed and bounded convex set with non-empty interior.
Let $F$ be a function that satisfies the following properties,
\begin{itemize}
\item
\emph{Valuation}: 
for any two convex bodies $B_i$ and $B_j$,
\begin{equation}
F(B_i \cup B_j) = F(B_i)+ F(B_j) - F(B_i \cap B_j).
\label{eq:association}
\end{equation}
\item
\emph{Invariance}: for any translation or rotation $g$, the measure $F$ is invariant, i.e, for any convex body $B$,
\begin{equation}
F(B) = F(gB).
\label{eq:tr}
\end{equation}
\item
\emph{Continuity}: for any convergent sequence of convex bodies, $B_i~\rightarrow~B$,
\begin{equation}
\lim_{i \rightarrow \infty} F(B_i) = F(B)
\end{equation}
\end{itemize}
Then for any $F$, there exist three constants $c_0$, $c_1$, and $c_2$ such that
\begin{equation}
F = \sum_{\nu=0,1,2} c_\nu \MF_\nu = c_0 \, A + c_1 \, L + c_2 \, \chi.
\end{equation}
where $\MF_\nu$ is $(A, L, \chi)$ for $\nu=(0,1,2)$, respectively.

\begin{figure}
    \hspace*{-0.5em}
    \subfloat[\label{fig:morphology:a}]{
        \includegraphics[width=0.24\textwidth]{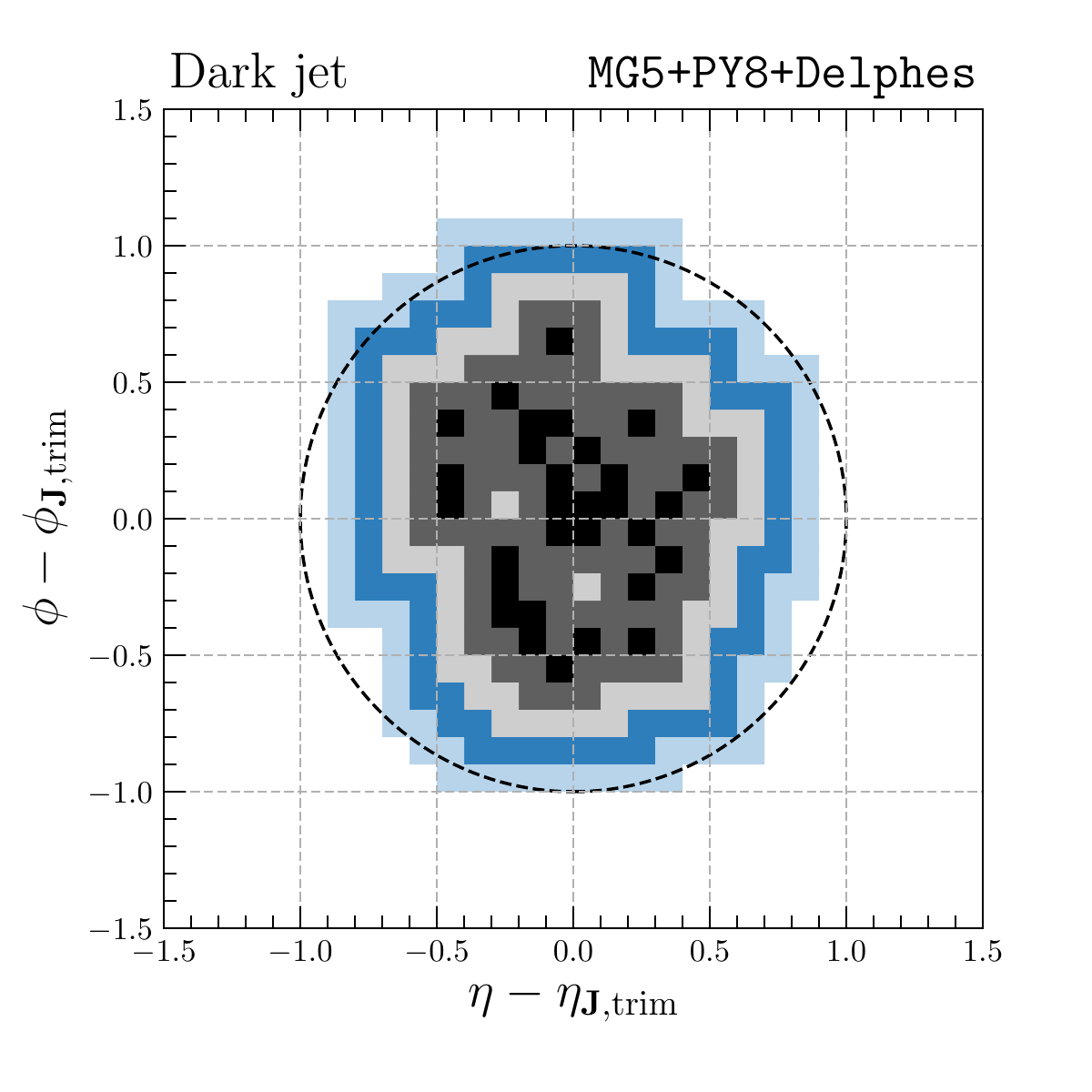} 
    }\hspace*{-1.5em}
    \subfloat[\label{fig:morphology:b}]{
        \includegraphics[width=0.24\textwidth]{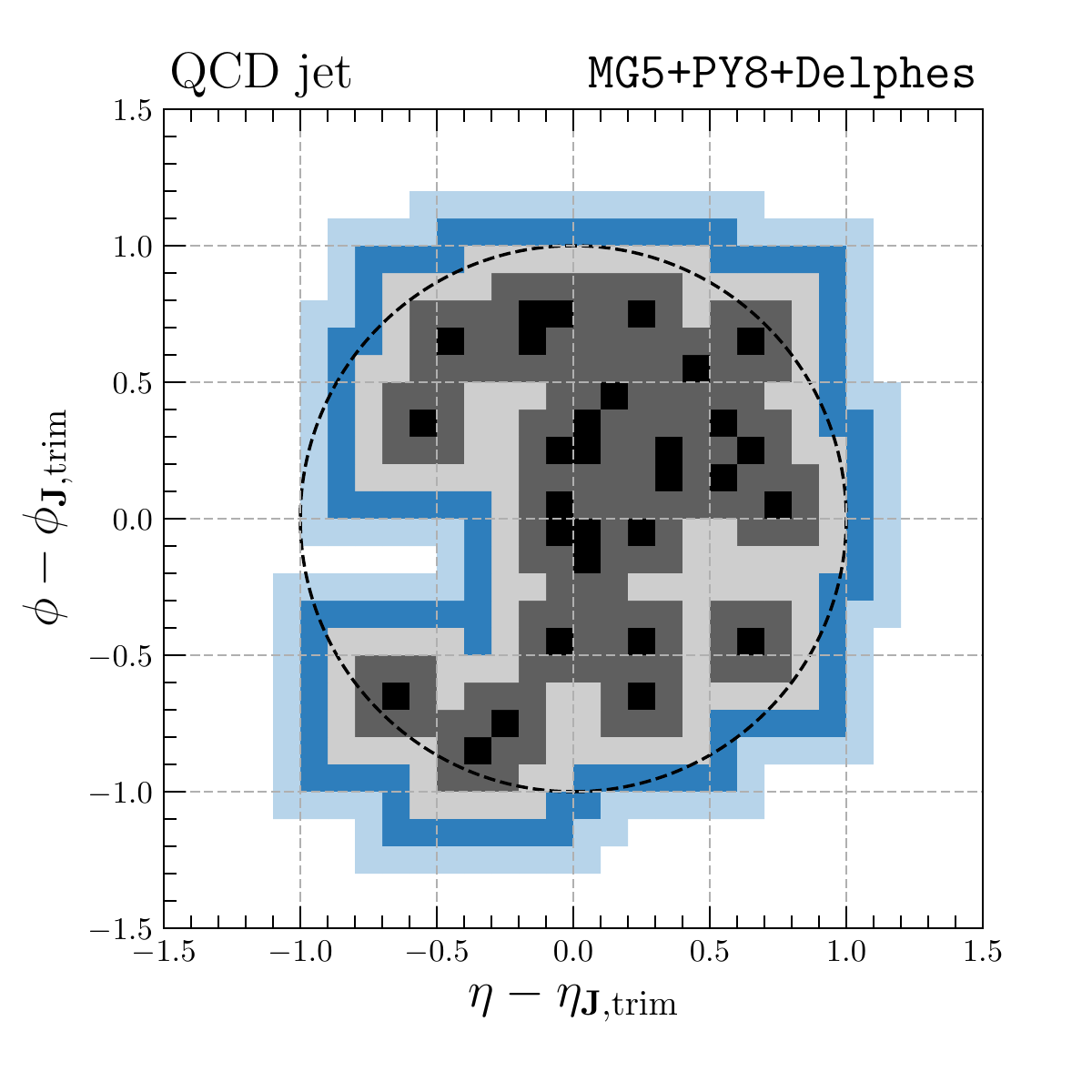}
    }
\caption{\label{fig:morphology}
Binary jet images of \subfigref{fig:morphology:a} a dark jet and \subfigref{fig:morphology:b} a QCD jet.  
Black dots are the active pixels in $P^{(0)}$ without any filtering.
Dark gray, gray, blue, and light blue pixels are pixels in $P^{(i)}-P^{(i-1)}$ for $i=1, 2, 3, 4$, respectively. 
Both of the binary images have $A^{(0)}=30$. 
The dark jet model is described in \sectionref{sec:benchmark_tagging}. 
}
\end{figure}

Hadwiger's theorem also holds in the geometry of the square lattice and pixelated image, but the context should be modified accordingly \cite{LEINSTER201281}. 
The geometry of the square lattice has a different distance function called the $L_1$ distance, which is a sum of the absolute value of the difference in each component as follows.
\begin{equation}
    ||\vec{R}_1 - \vec{R}_2||_1 = |\eta_1 - \eta_1| + |\phi_1 - \phi_1|.
\end{equation}
This distance is essentially identical to the length of the shortest path between two points on a square grid.
The points within unit $L_1$ distance from the origin is different to those in Euclidean geometry.
They form a square whose vertices are at $(0,1)$, $(0,-1)$, $(1,0)$, and $(-1, 0)$.

The statements of Hadwiger's theorem still holds under this geometry, but there are two modifications.
First, the invariance under translation and rotation is replaced by the isometry of the $L_1$ space.
The convexivity is replaced with $L_1$-convexivity. 
A given set $B$ is called $L_1$-convex if and only if 
there always exists a path connecting two points $\vec{R}_1$ and $\vec{R}_2$ in $B$,
and the components of the path are monotonic along the path.
One clear example illustrating the difference between those two convexivities is an L-shaped  region: it is not convex but $L_1$-convex.
After these modifications, we may safely use the MFs for the pixelated image analysis.

\subsection{Morphological Analysis on Jet Images}

The morphological analysis on jet images is then performed on the filtered distribution of jet constituents projected on the pseudorapidity-polar coordinate $(\eta,\phi)$.
We consider superlevel sets of the jet image, $P^{(0)}$, i.e., the set of pixels whose energy deposit $p_{T}^{(i,j)}$ is higher than the threshold value $p_T$ \cite{PhysRevE.53.4794},
\begin{equation}
    P^{(0)}[p_T] = \{(i,j) \, |\, p_{T}^{(i,j)} > p_T\},
\end{equation}
where $(i,j)$ is the integer coordinate of the given pixel.\footnote{The physical unit length of the grid is the hadronic calorimeter resolution $\Delta R = 0.1$ of our analysis.
The physical coordinates $(\eta, \phi)$ are obtained by multiplying $\Delta R$ to those integer coordinates.
}
The resulting binary images on a two-dimensional integer grid are used for the morphological analysis.
For the following discussion, we will omit the threshold argument $[p_T]$ unless it is required explicitly.

We then analyze the MFs of the images after dilation by a square called a structuring element to understand the geometric structure with the aid of mathematical morphology.
The dilation is useful for probing geometric features that are visible at the angular resolution of the size of the square.
For our pixellated image analysis, the structuring element $B^{(k)}$ is a square with side length $2 k+1$.
The dilated image $P^{(k)}$ is defined as follows.
\begin{eqnarray}
P^{(k)} 
& = & 
\{a + b \, \vert \, a\in P^{(0)}, b\in B^{(k)} \}, 
\\
B^{(k)} 
& = & 
\left\{(i,j) \, \vert \, i,j\in\{-k,-k+1,...,k-1,k\}\right\}.
\end{eqnarray}
Sample binary images are in \figref{fig:morphology}.
The binary image $P^{(k)}$ is analogous to a coarse-graining or smearing of the original binary image $P^{(0)}$.
We denote the three MFs of $P^{(k)}$ as $A^{(k)}$, $L^{(k)}$, and $\chi^{(k)}$. 
In \cite{Chakraborty:2020yfc}, we have shown that the MFs $ A^{(0)}$ and $A^{(1)}$ improve the top jet vs.~QCD jet classification.

We also note that the dilation by a square is good enough for retrieving the topology of an underlying smooth body where the point clouds are sampled.
The topology of the dilated image is sensitive to the structuring element in general, especially when we are using a finite number of samples. 
Still, the square is connected and sufficiently round so that the dilation by the square is a good topology estimation process without any glitches \cite{10.1145/1810959.1811015}.

We can get some intuitive idea of how the sequences of the MFs encode the geometric information of a given image by considering its limiting behavior.
For a scale $k$ much larger than the size of the image, $A^{(k)} \rightarrow (2k+1)^2$ because the details of the images are irrelevant to $P^{(k)}$.
In a different extreme case where $P^{(k)}$ is consisted by  $N$  sufficiently isolated clusters, the asymptotic behavior changes to $A^{(k)} \rightarrow N (2k+1)^2$.
Therefore, the sequence $A^{(k)}$ is sensitive to the number of clusters of active pixels in the jet image.

The intermediate behavior of the MF sequences $(A^{(k)}$,~$L^{(k)}$,~$\chi^{(k)})$ contains more details about the pixel distributions.
When $P^{(k)}$ is a convex body, the MFs of $P^{(k)}$ and $P^{(k+1)}$ satisfies the following recurrence relation \cite{LEINSTER201281},
\footnote{The equation can be derived from the theorem 6.2 of \cite{LEINSTER201281}, where the $L_1$-intrinsic volume $(V'_0,V'_1,V'_2) $ $= (\chi, L/2, A)$ and the scale factor $\lambda=2$}
\begin{eqnarray}
A^{(k+1)} & = &  A^{(k+1)}_{\mathrm{ext}} \equiv A^{(k)} + L^{(k)} + 4 \chi^{(k)}, \\
L^{(k+1)} & = &  L^{(k+1)}_{\mathrm{ext}} \equiv L^{(k)} + 8 \chi^{(k)}, \\
\chi^{(k+1)} & = & \chi^{(k+1)}_{\mathrm{ext}} \equiv \chi^{(k)}.
\label{eqn:recurrence}
\end{eqnarray}
The deviation from this relation signals that some change of the shape or topology occurs at the given angular scale. 
For example, if a hole or dent is completely filled during the dilation, the above recurrence relation is violated.
Therefore, the full sequences of the MFs contain useful information about the geometry of the binary image in general. 
The analysis is also a persistent analysis of geometric features of jet substructures, similar to \cite{Li:2020jdb}.

The recurrence relation also explains the asymptotic behavior of $A^{(k)}$. 
Suppose that the recurrence relations of the MFs hold after the given scale $k_0$.
The solution for $A^{(k+k_0)}$ in terms of the MFs of $P^{(k_0)}$ are as follows.
\begin{equation}
A^{(k_0+k)}_{\mathrm{ext}} = A^{(k_0)} + k \, L^{(k_0)} + 4 k^2 \chi^{(k_0)}
\end{equation}
For $k \gg k_0$, the area $A^{(k_0+k)}$ is approximately $4 k^2 \chi^{(k_0)}$, and the Euler 
characteristic $\chi^{(k_0)}$ can be interpreted as the number of clusters.

We now compare the area $A^{(k)}$ with the extrapolated area $A_{\mathrm{ext}}^{(k)}$ from the MFs of $P^{(k-1)}$ in order to check whether the dilation preserves the geometric features.
The difference $\Delta A^{(k)}$ is a useful measure for checking the geometric persistence,
\begin{eqnarray}
\Delta A^{(k)}
& = & 
A^{(k)}-A^{(k)}_{\mathrm{ext}}.
\end{eqnarray}

\begin{figure}[tb]
    \subfloat[\label{fig:independence:a}]{
        \includegraphics[width=0.235\textwidth]{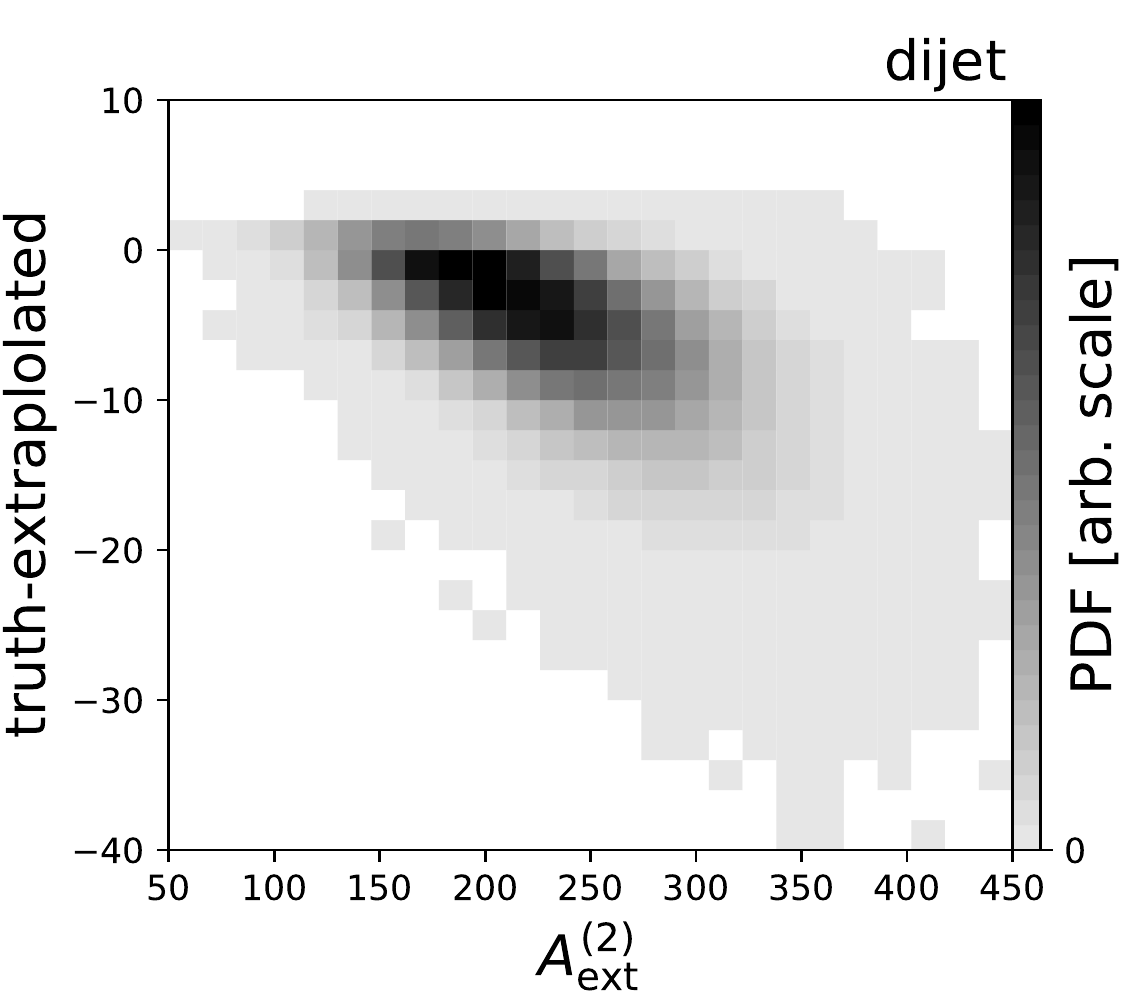}
    }\hspace*{-0.7em}
    \subfloat[\label{fig:independence:b}]{
        \includegraphics[width=0.235\textwidth]{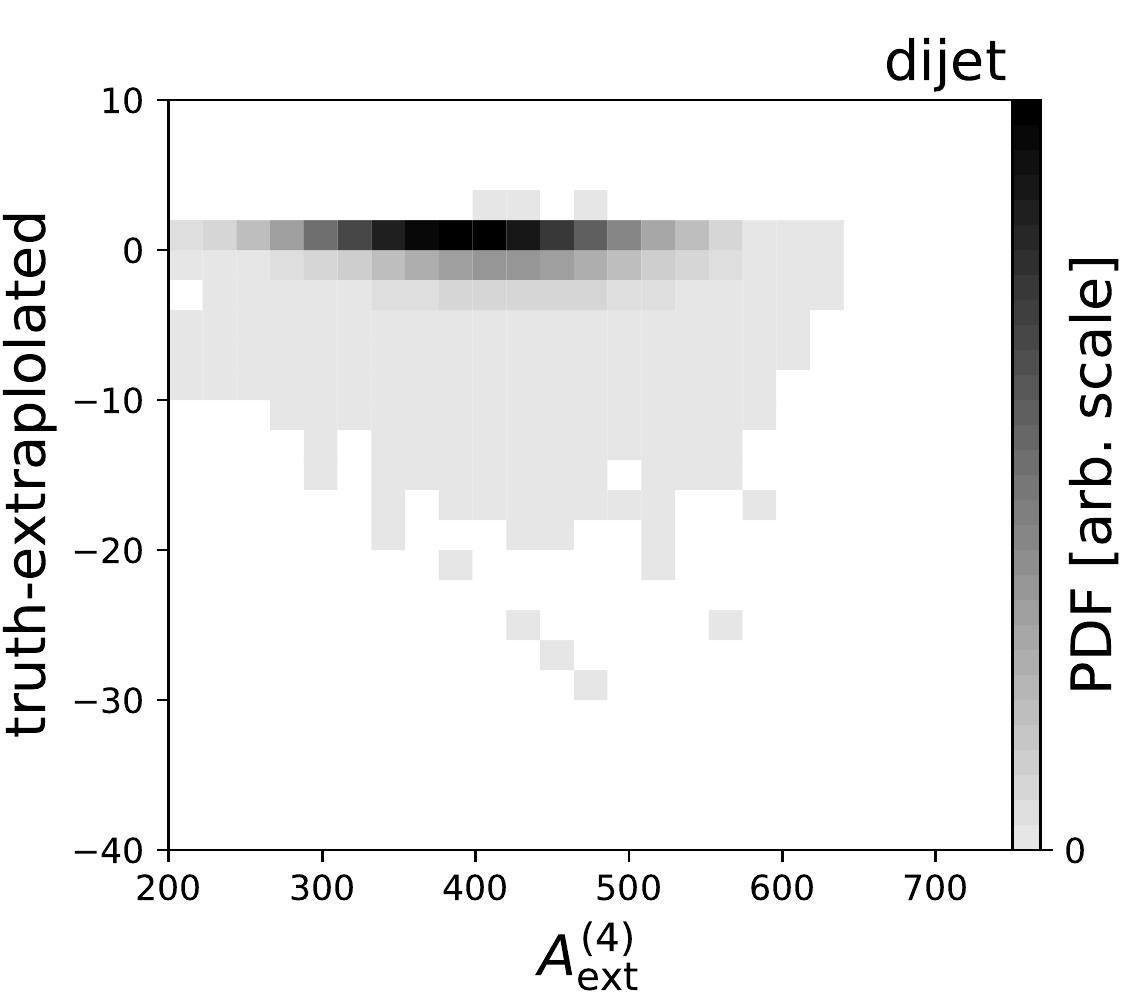}
    }
    
    \subfloat[\label{fig:independence:c}]{
        \includegraphics[width=0.235\textwidth]{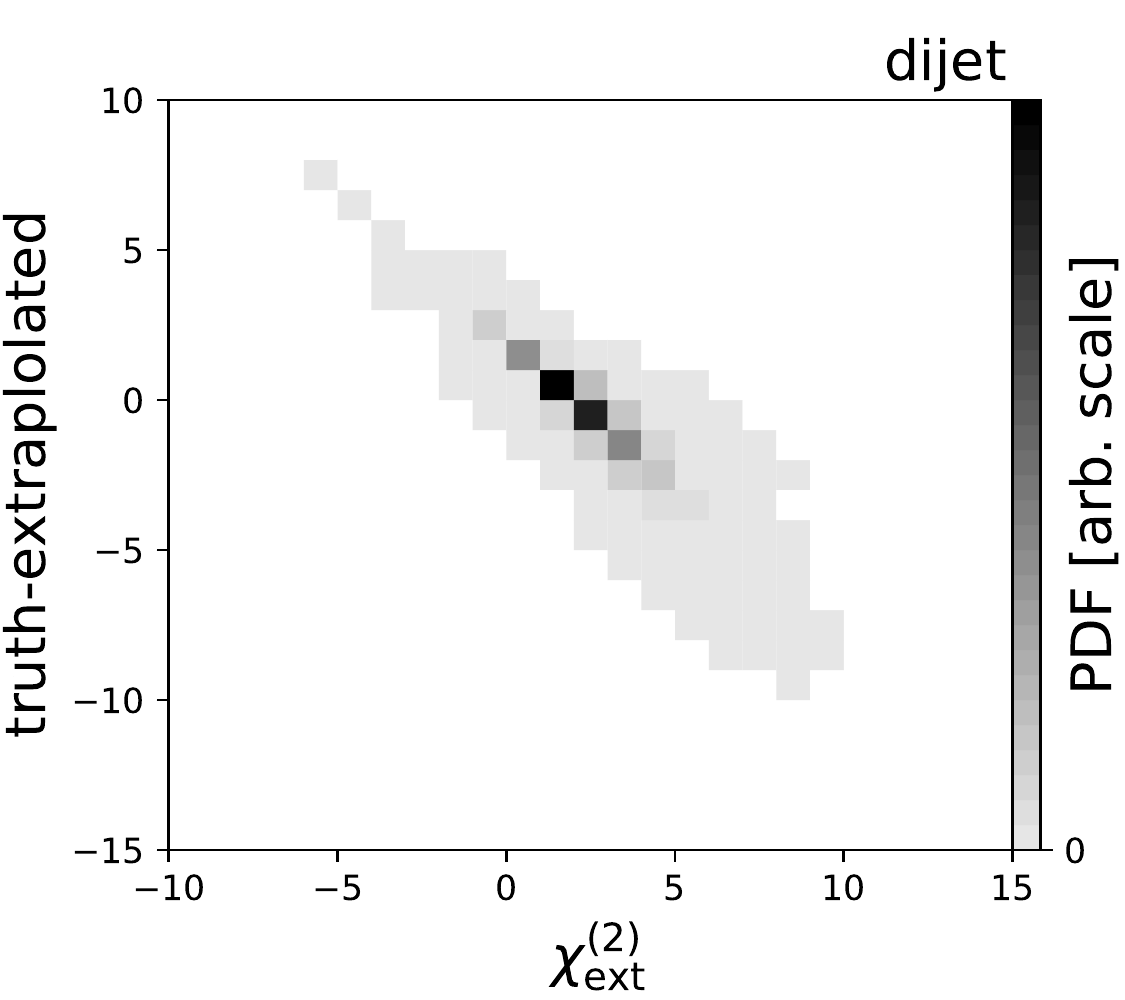}
    }\hspace*{-0.7em}
    \subfloat[\label{fig:independence:d}]{
        \includegraphics[width=0.235\textwidth]{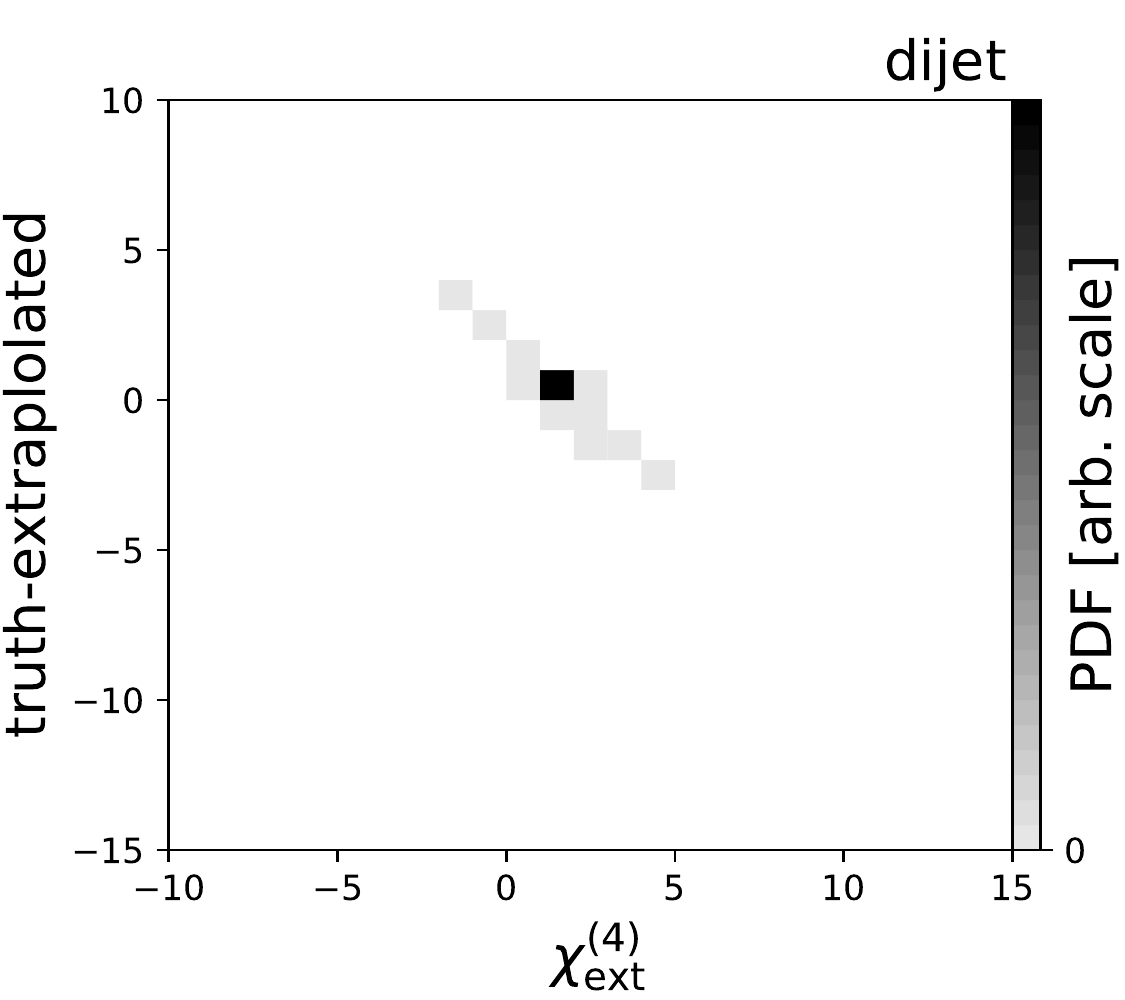}
    }
    \caption{
            The correlation between the MFs at a given scale $k$ and its extrapolated values from the scale $k-1$.
            The horizontal axis is the extrapolated value, and the vertical axis is the difference between the truth and extrapolated values.
            The upper plots (a) and (b) are for the area $A^{(k)}$, and the lower plots (c) and (d) are for the Euler characteristic $\chi^{(k)}$. 
            The left plots (a) and (c) are for $k=2$, and the right plots (b) and (d) are for $k=4$.
            For $k=4$, more samples have the difference zero since the dilation smooth out detailed features of jets and its geometry and topology becomes more and more trivial.
    }
    \label{fig:independence}
\end{figure}

\Figref{fig:independence} shows 2D histograms of $(A^{(k)}_{\mathrm{ext}},$  $ \Delta A^{(k)})$ of the leading $p_T$ jets of QCD dijet events with 
$p_{T,\jet}~$ $\in~[500,600]$~GeV.  
\Figref{fig:independence:a} for $k=2$ shows typical jets has lots of vibrant activities at the short scale so that the condition $\Delta A^{(k)} = 0$ can be easily violated for a small $k$.
A smeared image becomes more regular at the large scale, so that many of the samples has $\Delta A^{(k)} = 0$ as shown in \figref{fig:independence:b} for $k=4$.

A similar behavior can be directly seen in the Euler characteristics.
For a small $k$, the jets occasionally have subclusters, i.e., the intrinsic topology $\chi^{(k)}$ variate a lot. 
Therefore, the extrapolation $\chi_{\mathrm{ext}}^{(k)} = \chi^{(k-1)}$ is also quite different from $\chi^{(k)}$, {as shown in}
the 2D histogram of $(\chi^{(k-1)}, \chi^{(k)}-\chi^{(k-1)})$ in \figref{fig:independence:c}. 
For a large $k$, since we are analyzing a single jet, we expect that most of the events has $\chi^{(k-1)} \simeq \chi^{(k)} \simeq 1$ as in \figref{fig:independence:d}.
Note that $\chi^{(k)}-\chi^{(k-1)}$ is positive for some events, indicating that there are holes at the scale  $k-1$ and they are filled at the scale $k$.

Note that MFs are aggregated features, and their statistical fluctuations are smaller than the primitive inputs.
For example, the number of active pixels $A^{(0)}$ has fluctuation $\delta A^{(0)}/A^{(0)}\sim 1/\sqrt{A^{(0)}}$ but its pixel-by-pixel fluctuation is order 1.
As a result, the training of RN with MFs is potentially more stable against the fluctuation of the energy deposit of pixels, while CNN is more susceptible to that.

Neural networks trained on these MFs has useful geometric measures for solving the given task.
The MFs do not use energy weighting in contrast to other energy-weighted IRC safe jet substructure observables, so that all the jet constituents are treated equally once they pass the $p_T$ threshold.

\subsection{Convolution Representation of Minkowski Functionals}

The MFs are defined as an integral of local features in the continuum limit as in \eqref{eqn:mfs:continuum} so that they can be written as a sum of all the local contributions from finite-sized patches. 
This leads an interesting property of the MFs; they can be embedded in the CNN with finite-sized filters.

For example, the area of a two-dimensional region $S$ can be written as a following double integral of an indicator function $K$ of a square with side length $\ell$ and centered at $(0,0)$.
\begin{eqnarray}
    A & = &
    \int_S d^2\vec{r} \int_{\mathbb{R}^2} d^2 \vec{r}_0 \, \frac{1}{\ell^2} K_{\ell}(\vec{r}-\vec{r}_0) 
    \\
    K_{\ell}(x,y) & = & 
    \begin{cases}
    1 & x, y \in \left[-\frac{\ell}{2} , \frac{\ell}{2}\right] \\
    0 & \mathrm{otherwise}
    \end{cases}
\end{eqnarray}
By swap the order of the integration, we obtain the expression in the form of the sum of the local contribution of finite patches,
\begin{equation}
A = \int_{\mathbb{R}^2} d^2 \vec{r}_0 \left[ \int_S  d^2\vec{r} \, \frac{1}{\ell^2} K_{\ell}(\vec{r}-\vec{r}_0)  \right] 
\end{equation}
To discretize and evaluate this integral for the binary image on a square grid, the following marching square algorithm \cite{Goring:2013qya} is a fast and useful.

The marching square algorithm for square lattice process all the $2\times2$ subimages of given binary images and collect its local features for calculating the MFs.
The local features are summarized in \tableref{Table:lookup}.
Note that we {do not include} the boundary of the $2\times2$ subimages for this calculation.
\begin{itemize}
\item
For the area $A$, a subimage contribution is 1/4 of the number of its active pixels because a pixel belongs to four subimages.
\item
For the boundary length $L$, the contribution is local boundary length divided by 2 since every boundary belongs to two subimages.
\item
For the Euler characteristics $\chi$, we only need to count the number of outward corners, $N_\mathrm{out}$, and the number of inward corners, $N_\mathrm{in}$.
Since inward and outward corners have exterior angle $\pi/2$ and $-\pi/2$ respectively, the total curvature is just proportion to the difference between $N_\mathrm{out}$ and $N_\mathrm{in}$.
The Euler characteristic is then as follows,
\begin{equation}
    \chi 
    = 
    \frac{1}{2\pi} \left[\frac{\pi}{2} \left( N_\mathrm{out} - N_\mathrm{in} \right) \right]
    =
     \frac{1}{4} \left( N_\mathrm{out} - N_\mathrm{in} \right).
\end{equation}
Each corners are considered only once during the marching, 
the local contributions are $1/4$ for the outward corners and -$1/4$ for the inward corners.

Note that the Euler characteristic depends on the definition of the connectivity between two diagonally neighboring pixels. 
We define that pixels sharing a same vertex are connected, and the corresponding subimages have two inward corners.
\end{itemize}
For example, $(A, L, \chi)$ of an isolated pixel is the sum of 1, 2, 4, and 8 of the \tableref{Table:lookup}, and the value is $(1, 4, 1)$. 
This algorithm can be generalized for calculating MFs of images on other types of lattice, such as hexagonal pixels,
\footnote{
Note that the hexagonal grids are essentially identical to the plane of $\mathbb{R}^3$ with $L_1$ distance, with constraints ${x+y+z=0}$ \cite{413166, hexagon}.
The hexagonal pixels have rounder shape and larger symmetry groups than the square pixels, but its integral geometry is not trivial because of the projection.
Nevertheless, Hadwiger's theorem still holds in the $\mathbb{R}^3$, and the nontrivial $L_1$-intrinsic volumes $V_1'$ and $V_2'$ are proportional to the perimeter and area of the hexagonal pixels.
}
or for approximating MFs of raw images without pixelation \cite{Mantz_2008}.

\newcommand{\DrawMacroSquare}[4]{
\begin{tikzpicture}[baseline={([yshift=-.5ex]current bounding box.center)}]
\begin{scope}
\foreach \x in {0,...,2}
	\draw [gray] (-0.2+0.2*\x,-0.2) -- (-0.2+0.2*\x,0.2);
\foreach \y in {0,...,2}
	\draw [gray] (-0.2,-0.2+0.2*\y) -- (-0.2+0.4,-0.2+0.2*\y);
\ifthenelse{#1=1}{
\draw [fill=gray] (-0.2,0) rectangle ++(0.2,0.2);
}{}
\ifthenelse{#2=1}{
\draw [fill=gray] (0,0) rectangle ++(0.2,0.2);
}{}
\ifthenelse{#3=1}{
\draw [fill=gray] (-0.2,-0.2) rectangle ++(0.2,0.2);
}{}
\ifthenelse{#4=1}{
\draw [fill=gray] (0,-0.2) rectangle ++(0.2,0.2);
}{}
\end{scope}
\end{tikzpicture}
}

\begin{table}
\caption{\label{Table:lookup}The list of local contributions to the Minkowski functionals. 
We also show the corresponding $2\times2$ subimages for each argument.}
\begin{ruledtabular}
\begin{tabular}{ccccc|ccccc}
\multirow{2}{*}{arg.} & \multirow{2}{*}{conf.} & \multicolumn{3}{c|}{local contributions} &
\multirow{2}{*}{arg.} & \multirow{2}{*}{conf.} & \multicolumn{3}{c}{local contributions}
\\
\cline{3-5} \cline{8-10}
& & $A$ & $L$ & $\chi$ &
& & $A$ & $L$ & $\chi$ \\
\hline
0  & \DrawMacroSquare{0}{0}{0}{0} & \phantom{00}0\phantom{00}  & \phantom{00}0\phantom{00} & \phantom{00}0\phantom{00} &
8  & \DrawMacroSquare{0}{0}{0}{1} & 1/4  & 1   & 1/4 \\
1  & \DrawMacroSquare{1}{0}{0}{0} & 1/4  & 1   & 1/4   &
9  & \DrawMacroSquare{1}{0}{0}{1} & 1/2  & 2   & $-1/2$  \\
2  & \DrawMacroSquare{0}{1}{0}{0} & 1/4  & 1   & 1/4  &
10 & \DrawMacroSquare{0}{1}{0}{1} & 1/2  & 1   & 0    \\
3  & \DrawMacroSquare{1}{1}{0}{0} & 1/2  & 1   & 0    &
11 & \DrawMacroSquare{1}{1}{0}{1} & 3/4  & 1   & $-1/4$ \\
4  & \DrawMacroSquare{0}{0}{1}{0} & 1/4  & 1   & 1/4  &
12 & \DrawMacroSquare{0}{0}{1}{1} & 1/2  & 1   & 0   \\
5  & \DrawMacroSquare{1}{0}{1}{0} & 1/2  & 1   & 0  &
13 & \DrawMacroSquare{1}{0}{1}{1} & 3/4  & 1   & $-1/4$ \\
6  & \DrawMacroSquare{0}{1}{1}{0} & 1/2  & 2   & $-1/2$ &
14 & \DrawMacroSquare{0}{1}{1}{1} & 3/4  & 1   & $-1/4$ \\
7  & \DrawMacroSquare{1}{1}{1}{0} & 3/4  & 1   & $-1/4$ &
15 & \DrawMacroSquare{1}{1}{1}{1} & \phantom{00}1\phantom{00}  & \phantom{00}0\phantom{00} & \phantom{00}0\phantom{00} \\
\end{tabular}
\end{ruledtabular}
\end{table}

Since there are only 16 unique configurations for the $2\times2$ subimages, we may use the look-up table $\mathbf{v}^{k}(a)$, where $a=0,\cdots,15$ and $k\in\{A,L,\chi\}$, in \tableref{Table:lookup} for parameterizing the local contribution. 
{The MFs are then the sum of look-up table values as follows,}
\begin{equation}\label{eqn:filter}
(A^{(k)}, L^{(k)}, \chi^{(k)})
=
\sum_{i,j} \sum_{n,m \in \{0,1\}} \mathbf{v}\left(P_{(i+n)(j+m)}^{(k)}  f_{nm} \right) ,
\end{equation}
where $f_{nm} = ((1,2),(4,8))$, and $P_{ij}^{(k)}$ is 1 or 0 if $(i,j)$-th pixel of $P^{(k)}$ is active or not, respectively.

Note that all the steps for calculating MFs in this section can be written in terms of convolutions.
Let $p_T^{(i,j)}$ be the energy deposit of $(i,j)$-th pixel.
The calculation method of MFs discussed in this section can be summarized as follows.
\begin{eqnarray}
\nonumber
P^{(0)}_{ij}[p_T]
& = &
\theta( p_T^{(i,j)} - p_T)
\\*
\nonumber
P^{(k)} 
& = &
\theta( P^{(0)} * B^{(k)} )
\\*
\label{eqn:mfs:conv}
(A^{(k)}, L^{(k)}, \chi^{(k)})
& = &
\mathbf{v}(P^{(k)} * f)
\end{eqnarray}
where all the binary images in the above equations are considered as a function that gives 1 for active pixels and 0 for otherwise, $*$ is the discrete convolution.
The stacked convolution layers can simulate this algorithm, i.e., $B^{(k)}$ and $f$ can be considered as the weights of convolution layers, and the functions $\theta$ and $\mathbf{v}$ can be modelled by $1\times1$ convolutions \cite{DBLP:journals/corr/LinCY13}.
Therefore, $A^{(k)}$, $L^{(k)}$, and $\chi^{(k)}$ are in principle covered by a CNN trained on jet images.

One subtle point is that this closed expression contains a step function, which has a point of discontinuity.
The CNN with a finite number of filters and smooth activation functions may have difficulty on accessing this variable set since the network itself is a smooth function.  
A similar situation may happen on the CNN with $L_2$ regularizers. 
We will show an example that the tagging performance of the CNN is improved by adding MFs to the inputs.


\section{Energy Correlator based Neural Networks for Jet Substructure}
\label{sec:irc_safe_net}

The energy dependence of MFs in \eqref{eqn:mfs:conv} is nonlinear, while
many theory-motivated jet substructure variables typically have a multilinear energy dependence; these types of variables are called IRC safe energy correlators \cite{Tkachov:1995kk, Komiske:2017aww}.
Since the counting variables complement those variables, 
we may use a neural network model representing the IRC-safe energy correlators and provide the MFs as additional inputs.
In this section, we briefly review two examples: the IRC-safe relation network \cite{Lim:2018toa,Chakraborty:2019imr,Chakraborty:2020yfc}, and the energy flow network \cite{Komiske:2018cqr}

\subsection{Relation Network}

The relation network (RN) is mainly designed for capturing the common properties of relational reasoning.
For example, if we use the momentum $p_i$ of the $i$-th constituents of the jet as a network input, we can build one simplest model of RN with two scalar functions $f$ and $g$ as follows,
\begin{equation}
\label{eqn:rn:simplest}
    f \left[ \sum_{i\in a, j\in b} g(p_i, p_j) \right],
\end{equation}
where $a$ and $b$ are labels for subsets of jet constituents.
If we impose the IRC-safe constraints \cite{Tkachov:1995kk, Komiske:2017aww}, the function $g$ should be bilinear in the constituent $p_T$ and the coefficients $\Phi_{ab}$ should depend only on the relative angular distance between the jet constituents, $R_{ij}$.
The following is then the basic form of the IRC-safe RN for the jet substructure,
\begin{equation}
\label{eqn:rn:irc_safe}
    f \left[ \sum_{i\in a, j\in b} p_{T,i} p_{T,j} \Phi_{ab}(R_{ij}) \right].
\end{equation}
The summation in the above equation is a nested loop over the jet constituents. 
Nevertheless, this part can be  simplified to a single summation as we describe below.

We introduce the following two-point energy correlation $S_{2,ab}$ that accumulates energy correlations at a given angular scale $R$.
\begin{equation}
\label{eqn:s2}
S_{2, ab}(R)= \sum_{i\in a, j\in b}  p_{T,i} p_{T,j} \delta(R-R_{ij}).
\end{equation}
By using $S_{2,ab}$, the nested summation in \eqref{eqn:rn:irc_safe} can be replaced to a single integral as follows,
\begin{equation}
\label{eqn:s2:rn}
    \int dR \,S_{2,ab}(R) \Phi_{ab}(R).
\end{equation}
This model covers various jet substructure variables.
For example, the two-point energy correlation functions {EFP}$^n_2$ \cite{Larkoski:2013eya,Komiske:2017aww} can be written in terms of a linear combination of the $S_2$ as follows,
\begin{equation}
{\rm EFP}^{n}_{2,ab} = \int^{\infty}_{0} dR \, 
S_{2, ab}(R)\, R^{n},   
\end{equation}
Therefore, this network covers all information encoded in $\mathrm{EFP}^n_2$.

For the practical use of this RN with IRC-safe constraints, we discretize the integral in \eqref{eqn:s2:rn} by binning the integrand with bin size $\Delta R$.
The discrete version of $S_{2,ab}$ is then defined as follows.
\begin{equation}
S_{2,ab}^{(k)} = \int_{ k\Delta R}^{ (k+1)\Delta R } dR \, S_{2,ab}(R),
\end{equation} 
where $k$ is the bin index.
The integral in \eqref{eqn:s2:rn} can be expressed as a inner product between $S_{2,ab}^{(k)}$ and a weight vector $\Phi_{ab}^{(k)}$,
\begin{equation}
    \int dR \,S_{2,ab}(R) \Phi(R) = \sum_k S_{2,ab}^{(k)} \Phi_{ab}^{(k)}.
\end{equation}
For our numerical study, we take bin size $\Delta R=0.1$, which is the hadronic calorimeter resolution. 
The $S_2$'s are directly calculated from the HCAL and ECAL outputs.  
If we use an MLP to model the function $f$ of the RN in \eqref{eqn:rn:irc_safe}, we can embed $\Phi^{(k)}$ to the first fully-connected layer.
The fully-connected layer that maps one input $\sum_k S_{2,ab}^{(k)} \Phi_{ab}^{(k)}$ to the latent dimension is equivalent to a fully connected layer that maps $S_{2,ab}^{(k)}$'s to the latent dimension, i.e.,
\begin{equation}
W_{l} \sum_k S_{2,ab}^{(k)} \Phi_{ab}^{(k)} = \sum_k W_{lk} S_{2,ab}^{(k)},
\quad
W_{lk} = W_{l} \Phi_{ab}^{(k)}.
\end{equation}
The RN is modelled by an MLP taking $S_{2,ab}^{(k)}$, and the first layer can be regarded as \emph{a trainable two-point energy correlation}.

\subsection{Energy Flow Network}
Energy flow network (EFN) \cite{Komiske:2018cqr} is also a graph neural network based on the energy correlators, but this network uses only pointwise features. 
This network is based on the deep set architecture \cite{NIPS2017_f22e4747}, i.e.,
\begin{equation}
    f\left[\sum_{i\in a} g (p_i) \right].
\end{equation}
As discussed before, this pointwise feature $g(p_i)$ should be a linear function of energy when the IRC-safe constraint is assumed, and we have the following model of the EFN.
\begin{equation}
    f\left[\sum_{i\in a} p_{T,i} \Phi (\vec{R}_i) \right]
\end{equation}
For the pixelated image analysis, 
the $p_T$-weighted sum over the jet constituents is replaced to the energy-weighted sum over all pixels,
\begin{equation}
    \sum_{i\in a} p_{T,i} \Phi (\vec{R}_i) \approx \sum_{i,j} P_{T}^{(ij)} \Phi_{ij},
\end{equation}
where $P_T^{ij}$ is the energy deposit of the $(i,j)$-th pixel, and $\Phi_{ij}$ is the corresponding angular weights.

When we replace $f$ with an MLP, the angular weights $\Phi_{ij}$ can be absorbed into the MLP.
The product between the weights $W_\ell$  of the first dense layer and $\Phi_{ij}$ can be considered as an effective weights $W_{\ell ij}$ of an MLP taking $P_{T}^{(ij)}$ as inputs, i.e., the dense layer can be rewritten as follows.
\begin{equation}
W_{\ell} \left[ \sum_{i,j} P_{T}^{(ij)} \Phi_{ij} \right] = \sum_{i,j} P_T^{(ij)} W_{\ell ij}, \quad W_{\ell ij} = W_{\ell} \Phi_{ij}.
\end{equation}
Therefore, an MLP for the pixelated image analysis models the EFN for the pixelated jet image.

Note that using the standardized inputs results does not change the conclusion since the standardization is a linear transformations. 
Let us consider the following transformation of the inputs and parameters of the dense layer transforms,
\begin{eqnarray}
    P^{(ij)}_T
    & \rightarrow & 
    \frac{P^{(ij)}_T - \mu^{(ij)}}{\sigma^{(ij)}},
    \\
    W_{\ell ij} 
    & \rightarrow & 
    \sigma^{(ij)} W_{\ell ij},
    \\
    B_{\ell} 
    & \rightarrow & 
    \sum_{i,j} \mu^{(ij)}  W_{\ell ij} + B_{\ell},
\end{eqnarray}
where $\mu^{(ij)}$ and $\sigma^{(ij)}$ are the mean and standard deviation of the inputs.\footnote{For the pixels which do not have energy variations, we assign $\sigma^{(ij)} = 1$.}
The first dense layer, $ \sum_{i,j} P^{(ij)}_T W_{\ell ij}  + B_{\ell}$ is invariant under this transformation, we may safely use the MLP for the standardized image to model the EFN.

\section{Combined Network Setup}
\label{sec:combined_network}
In this section, we describe the network that combines the morphological analysis and  the RN or EFN.

\subsection{Network Inputs}

For the morphological analysis, we use the MFs up to $k=6$ 
and denote them as $x_{\mathrm{morph}}$, 
\begin{equation}
\xmorph = \bigcup_{p_T\,\mathrm{threshold}}\{ 
A^{(k)}, L^{(k)}, \chi^{(k)} \, | \,
k=0,\cdots,6
\}.
\end{equation}
We use the following $p_T$ thresholds: default threshold of the detector simulation\footnote{0.5 GeV for the electronic calorimeters and 1.0 GeV for the hadronic calorimeters. This filtering is performed before the pixellation.}, 2, 4, and 8 GeV.

For the IRC-safe relation network, we used the two-point energy correlation $S_{2,ab}$ of the following subsets of jet constituents. 
\begin{itemize}
\item the trimmed jet $\jet_{\trim}$ \cite{Krohn:2009th}, denoted by $h$, 
\item the compliment set of $\jet_{\trim}$,   denoted by $s$, 
\item the leading $p_T$ subjet $\jet_1$,   denoted by $1$,
\item the compliment set of $\jet_1$,  denoted by $c$.  
\end{itemize}
Using these subsets is effective in the top tagging \cite{Chakraborty:2020yfc}.
We use the following sets of binned two-point correlations as inputs of the RN, 
\begin{eqnarray}
\nonumber
\xtrim
& = &
\{ S^{(k)}_{2, hh},  S^{(k)}_{2, {\rm soft}} \equiv  2 S^{(k)}_{2, hs} + S^{(k)}_{2, ss}\, | \, k=0,\cdots,14 \},
\\
\nonumber
\xleadingjet
& = &
\{ S_{2,11}^{(k)} \, | \, k=0,1,2 \} \cup 
\{ S_{2,1c}^{(k)} \, | \, k=0,\cdots,9 \}
\\
& &
\phantom{00} \cup 
\{ S_{2,cc}^{(k)} \, | \, k=0,\cdots,14 \},
\end{eqnarray}

In addition to those MFs and two-point energy correlations,
we provide $p_T$ and mass for each jet, trimmed jet, and leading $p_T$ subjets as additional inputs to give information regarding jet kinematics, and we denote them as $\xkin$.

\begin{equation}
    \xkin= \{ p_{T,\jet}, m_{\jet}, p_{T,\jet_{\trim}}, m_{\jet_{\trim}} , p_{T,\jet_1}, m_{\jet_1} \}.
\end{equation}

\subsection{Network Architecture}

We use the following setup to transform the given inputs to the desired outputs for the binary classification.
We first use MLPs to encode each of the primitive inputs $\xmorph$, $\xtrim$, and $\xleadingjet$ into latent spaces of dimension 5,
\begin{eqnarray}
h_{\mathrm{morph}}
&=& 
\mathrm{MLP}_{\mathrm{morph}}(x_{\mathrm{morph}}, x_{\mathrm{kin}}),
\\
h_{\trim}
&=& 
\mathrm{MLP}_{\trim}(x_{\trim}, x_{\mathrm{kin}}),
\\
h_{\jet_1}
&=& 
\mathrm{MLP}_{\jet_1}(x_{\jet_1}, x_{\mathrm{kin}}).
\end{eqnarray}
All the MLPs used in this section take the kinematic inputs $\xkin$ as additional inputs.
Those latent space features are mapped into the {classifier outputs} $\hat{y}$ the by another MLP,
\begin{equation}
    \logit(\hat{y}) = \mathrm{MLP}_{\mathrm{out}} (h_{\mathrm{morph}}, h_{\trim}, h_{\jet_1}, x_{\mathrm{kin}}),
\end{equation}
where $\logit(\hat{y})$ is the inverse of the standard logistic function, $\log(\hat{y}) - \log(1-\hat{y})$. 
For the analysis using only the subset of the inputs, we take only the relevant latent space features.
We denote this setup as RN+MF, and the pure RN setup without morphological analysis as RN.

We will use this network for binary classifications, trained by minimizing the binary cross-entropy loss function. 
\begin{equation}
    \label{eqn:loss_ce}
    \mathcal{L}_{\mathrm{CE}} 
    =
    -\frac{1}{2} \E\left( \log \hat{y} \,|\, y=1 \right)
    - \frac{1}{2} \E\left( \log (1-\hat{y}) \,|\, y=0 \right),
\end{equation}
where $y=1$ indicates the signal samples, and $y=0$ indicates the background samples.
The priors for each class is $1/2$.
All the hidden layer's weights are L2 regularized with a {weight decay coefficient} of 0.001.
The network is trained by ADAM optimizer \cite{ADAM} with default parameters, and we adopt the temporal exponential moving average on trainable parameters after ignoring the early 50 epochs.
The ratio between training, validation, and test datasets is 9:1:10.
We stop training when the validation loss does not improve for 50 epochs.
We iterate this procedure for different numbers of minibatches of  20, 50, 100, and 200,
and choose the results with the largest validation AUC.
All of these setups are implemented using \texttt{Keras} \cite{chollet2015keras} with \texttt{TensorFlow} backend \cite{tensorflow2015-whitepaper}.
Finally, all  inputs are standardized, and we also reweight events to make 
the $p_T$ distribution flat in order to marginalize learning  from $p_{T,\jet}$ distribution.

We also remark that in a limit of large width of the MLPs and small bin size for $S_2$ and MFs, this network setup corresponds to the following smooth model,
\begin{eqnarray}
\nonumber
h_{\mathrm{MA}} & = & \Psi_{\mathrm{MA}} \left[ \int_0^\infty d p_T \int_0^\infty dR \, \mathrm{MF}_j(R; p_T) \Phi_j(R; p_T) ; x_{\mathrm{kin}} \right]
\\
\nonumber
h_{\mathrm{RN}} & = &
\Psi_{\mathrm{RN}} \left[ \sum_{a,b} \int_0^\infty d R\,  S_{2,ab}(R) \Phi(R) ; x_{\mathrm{kin}} \right]
\\
\hat{y}
& = &
\Psi_{\mathrm{out}} \left[ h_{\mathrm{MA}}, h_{\mathrm{RN}}; x_{\mathrm{kin}} \right],
\end{eqnarray}
where all the $\Phi$ and $\Psi$ are some scalar functions.
This expression can help discuss the relationship between the morphological analysis and other networks working on the momenta of jet constituents without pixelation, such as ParticleNet \cite{Qu:2019gqs}. However, the discussion is beyond the scope of this paper.

\subsection{Convolutional Neural Network and Energy Flow Network}
We compare this RN+MF to the following CNN and EFN.

Our baseline CNN is trained on the preprocessed jet images, as described in \cite{Chakraborty:2020yfc}.
\begin{enumerate}
    \item The jet constituents are reclustered by $k_T$ algorithm \cite{Catani:1993hr,Ellis:1993tq} with radius parameter 0.2.
    \item 
    Set the center of $(\eta,\phi)$ coordinate to be the leading $p_T$ subjet axis.
    \item
    Rotate $(\eta,\phi)$ plane about the origin so that the subleading $p_T$ subjet is on the positive $y$ axis.
    \item
    If the third leading $p_T$ subjet exists and has negative $x$ value, 
    flip the $x$ axis so that the third subjet is always on the right side of the image.
    \item
    Pixelate the jet constituents to get the jet image.
\end{enumerate}
The preprocessed jet image is a two-dimensional $p_T$ weighted histogram of jet constituents on a range $[-1.5, 1.5] \times [-1.5, 1.5]$ with bin size $0.1\times 0.1$.
We denote the set of energy deposits for each pixels as follows,
\begin{equation}
    x_\mathrm{image} = \{ P_T^{(ij)}\,|\, i,j=-15,\cdots,14 \}.
\end{equation}
The image input $x_\mathrm{image}$ is provided to networks after standardization. 
In summary, the preprocessed images are aware of the most energetic subjet locations, and the relative position of the two subleading $p_T$ subjets.

The CNN consists of six convolutional layers. 
The filter size is $3\times 3$, and a pooling layer with pool size $2\times 2$ is inserted for every three convolutional layers.
After then, all the spatial dimensions are flattened, and a $1\times1$ convolution maps the intermediate outputs to latent space with dimension 10.\footnote{We have checked the classification performance of the CNNs with the latent dimensions 5, 10, 20, and 100, and 10 was the best.}
These latent space features are then concatenated to the kinematic inputs $x_{\mathrm{kin}}$, and we use an MLP to transform them into the desired classifier output.
The training setups are the same as RN+MF, but we scan by minibatch numbers 100, 200, and 500.

Although CNN can represent MFs, we may explicitly provide the MFs to the CNN.
As discussed earlier, CNN may experience technical difficulty expressing MFs through the training because the MFs are not smooth functions of the jet image. 
We additionally consider a CNN 
whose MLP at the end receives $h_{\mathrm{morph}}$ as additional latent space inputs.
We denote this setup as CNN+MF.

We model the pointwise correlation of the EFN by an MLP  {with three} hidden layers and 10 outputs.
The {first} hidden layer has 50 (200) outputs, while the others have 200 outputs.
The input image is concatenated with 
$\xkin$.
The outputs are then provided to another MLP that converts those inputs to the classifier, similar to that of the CNN.

\Tableref{tableii} lists the combination of inputs we study in this paper, and training costs for the classification problems that is discussed in \sectionref{sec:training_performance}. 
Some notable differences between the inputs to the CNNs and the RN+MFs are as follows.

The baseline CNN takes a large number of inputs since they are taking the whole image.
However, the {detector hits are} sparsely distributed over the images since the center of the images contains more information while the outer region of the jet image has sparse soft activities.
The CNN has to distill the useful information from this sparse dataset.
On the other hand, RN only takes the basis for the two-point energy correlators.
The soft activities are collected to each bin of $S_2$, and the resulting number of inputs is only O[100].

The number of MF inputs is $3\times7$ for each binary image given energy thresholds.
This is also a relatively small number compared to the dimension of the image inputs.
We also note that as $k$ increases, the change in geometry of the dilated image $P^{(k)}$ becomes more regular, and the MFs are getting dependent on their previous values in the sequences.
The cutoff for $k$ may be fine-tuned further, but we use 7, which effectively smoothes out geometric features below the angular scale of 1.5.
The latter terms in the sequence merely validate the regularity in 
dilation, and dropping some of them may not affect the performance significantly.

\begin{table}
\caption{
The number of inputs $N_{\mathrm{input}}$, the number of trainable parameters $N_{\mathrm{param}}$.
The number of inputs includes dummy inputs since each $S_2$'s are saved on length 20 vectors.
For EFNs, the number of params in parenthesis is the number for reduced setup with 50 energy correlators while the nominal setup has 200 energy correlators.
}
\begin{ruledtabular}
\begin{tabular}{llcc}
& 
inputs & 
$N_{\mathrm{input}}$ & 
$N_{\mathrm{param}}$ 
\\
\hline
  MF & $\xmorph$, $\xkin$ & \phantom{0}90 & 102,407 \\
  RN & $\xtrim$, $\xleadingjet$, $\xkin$ & 106 & 149,212 \\
   RN+MF & $\xmorph$, $\xtrim$, $\xleadingjet$, $\xkin$ & 190 & 209,617 \\
  CNN & $\ximage$, $\xkin$ & 906 & 131,740  \\
  CNN+MF & $\ximage$, $\xmorph$, $\xkin$  & 990 & 228,235 \\
   \multirow{2}{*}{EFN} & \multirow{2}{*}{$\ximage$, $\xkin$} & \multirow{2}{*}{906} & 202,167 \\
  & & & (141,762) \\ 
  \multirow{2}{*}{EFN+MF} & \multirow{2}{*}{$\ximage$, $\xmorph$, $\xkin$}  & \multirow{2}{*}{990} & 408,417 \\
  & & & (348,012) \\
\end{tabular}
\end{ruledtabular}\label{tableii}
\end{table}

\section{Jet Tagging Performance Comparison }
\label{sec:benchmark_tagging}

\subsection{Semi-visible Jet Tagging}
\label{sec:benchmark_tagging:semi_visible}
\begin{figure*}
    \subfloat[\label{fig:ms:a}]{
        \includegraphics[width=0.45\textwidth]{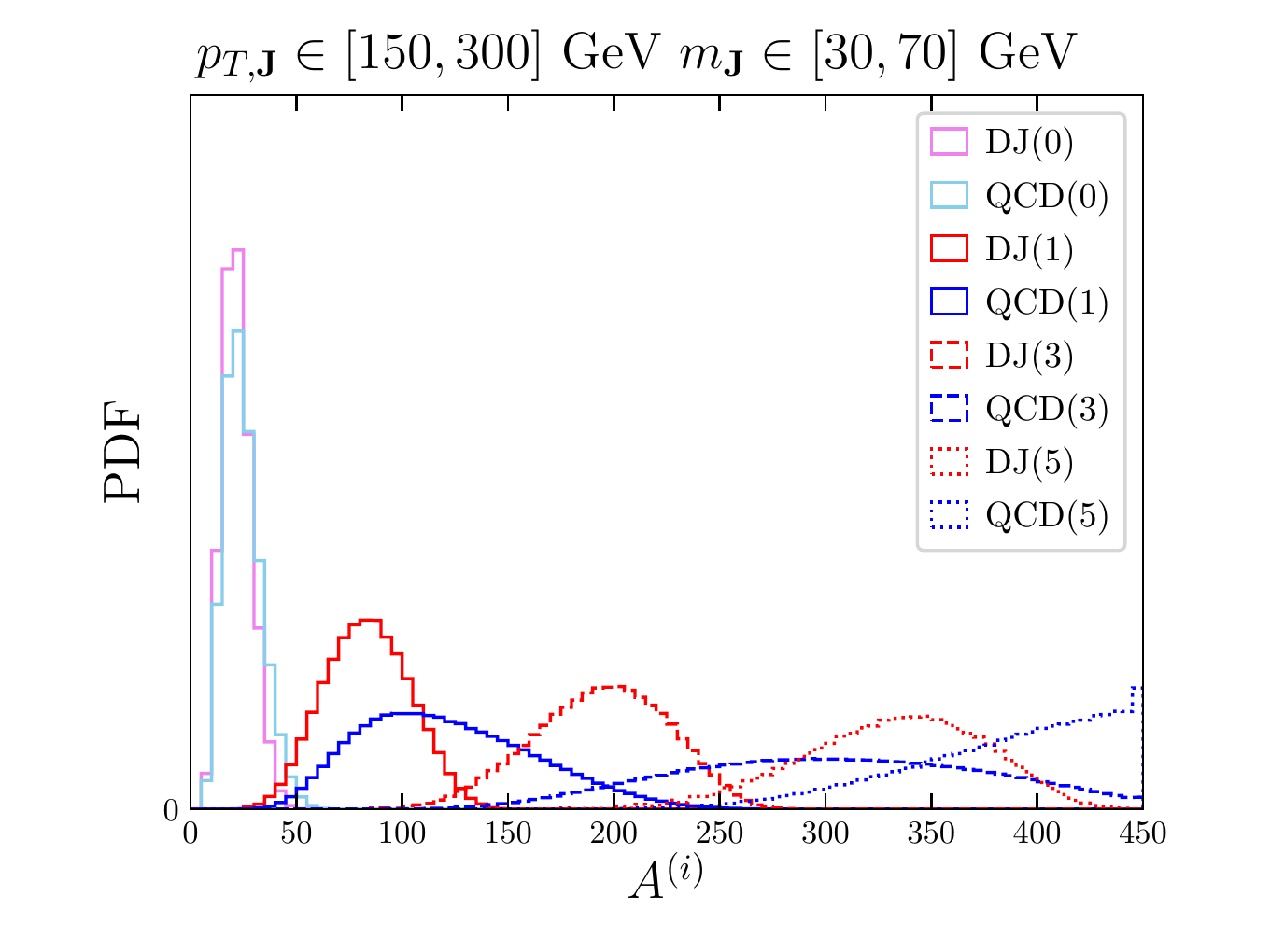}
    }
    \subfloat[\label{fig:ms:b}]{
        \includegraphics[width=0.45\textwidth]{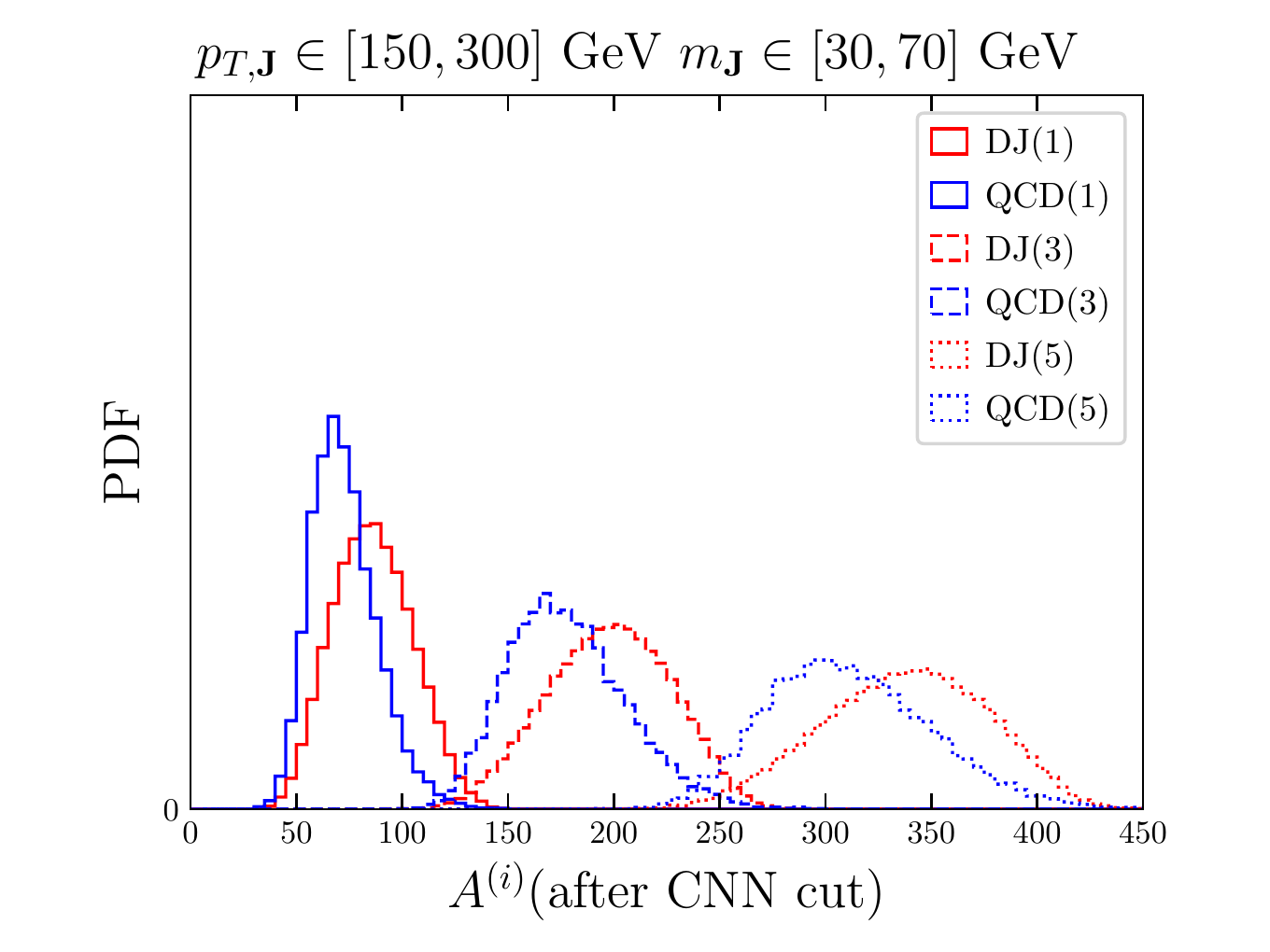}
    }
\caption{
\label{fig:ms} 
Left: distributions of MFs:
$A^{(0)}$ (light color), $A^{(1)}$ (solid), $A^{(3)}$ (dashed), and  $A^{(5)}$ (dotted) of dark jets (red)
and QCD jets (blue).
We select leading $p_T$ jets with $p_{T,\mathbf{J}} \in [ 150, 300]\,\mathrm{GeV}$,
and $m_{\mathbf{J}} \in [30,70]\,\mathrm{GeV}$.  
Right: The distribution of MFs after rejecting 10\% signal events by the CNN.  
1.5\% of QCD events remain after the selection.   
}
\end{figure*}

\begin{figure}
\includegraphics[width=0.45\textwidth]{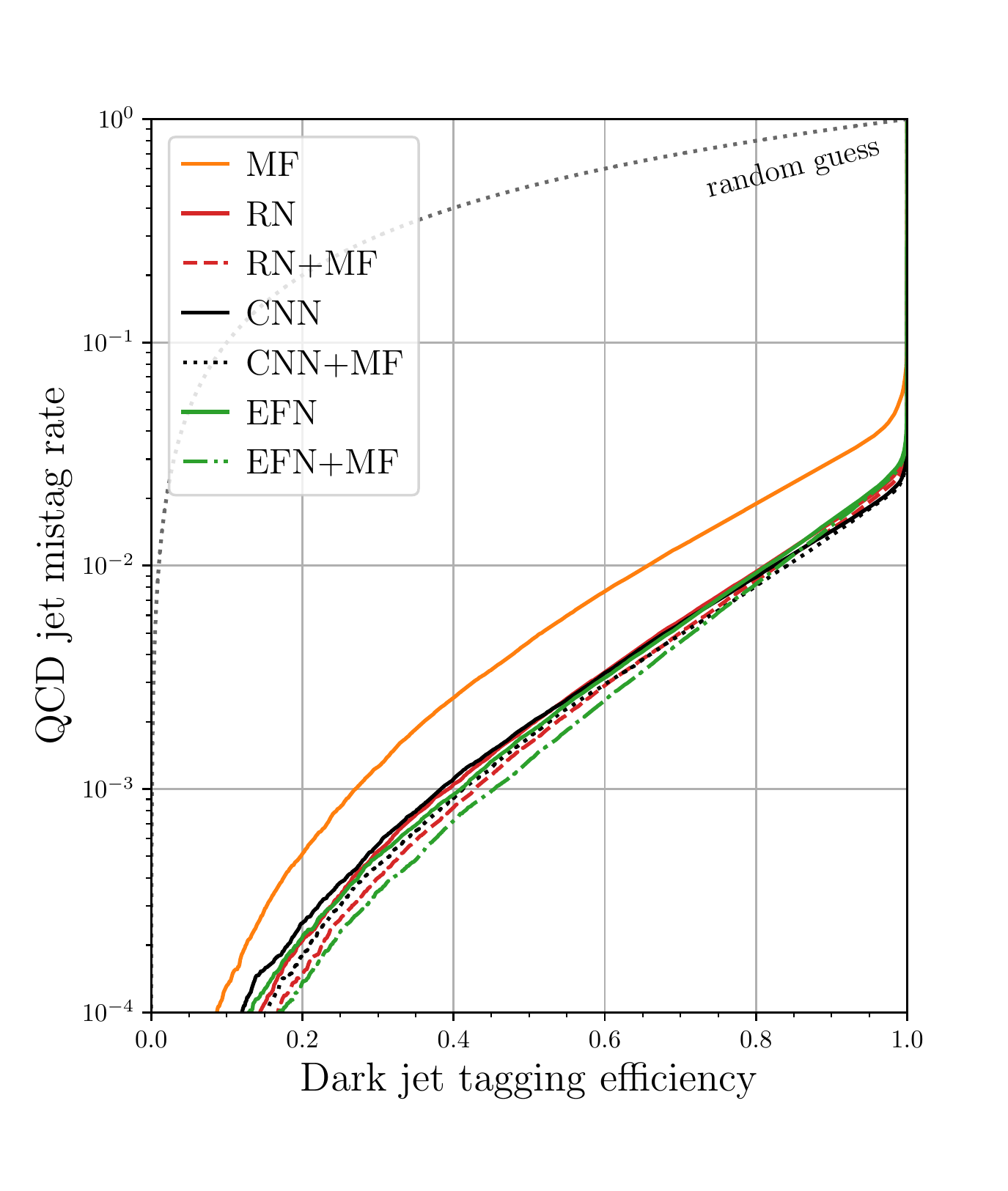}
\caption{\label{fig:roc:darkjet}
ROC curves of various classification models for dark jets vs.~QCD jets.
}
\end{figure}

\begin{table}
\caption{
\label{table:auc:dark_jet}
AUCs of various dark jet taggers.
The EFN models have 200 hidden features at the first dense layer.
We also show the training time $t_{\mathrm{train}}$ and the number of epochs at the end of the training, $N_{\mathrm{train}}$ {for} mini-batch numbers $N_{\mathrm{batch}}=$ 20 and 200.
}
\begin{ruledtabular}
\begin{tabular}{lccc}
      & 
      \multirow{2}{*}{AUC}
      &
      \multicolumn{2}{c}{$t_{\mathrm{train}} / N_{\mathrm{epoch}}$}
\\
& & $N_{\mathrm{batch}}=20$ & $N_{\mathrm{batch}}=200$ \\
\hline 
MF      & 0.9897   & \phantom{0}793 s / 564 epochs & \phantom{00}954 s / 363 epochs \\
\hline 
RN      & 0.9950   & \phantom{0}929 s / 434 epochs & \phantom{0}2468 s / 560 epochs\\
RN+MF   & 0.9955   & 1128 s / 429 epochs & \phantom{0}2288 s / 556 epochs\\
CNN     & 0.9953   & & 11401 s / 327 epochs \\
CNN+MF  & 0.9956   & & 19610 s / 543 epochs \\
\hline
EFN     & 0.9950   & 2222 s / 220 epochs & \phantom{0}2141 s / 163 epochs \\
EFN+MF  & 0.9955   & 1988 s / 190 epochs & \phantom{0}2270 s / 172 epochs \\
\end{tabular}
\end{ruledtabular}\label{tableiii}
\end{table}

\begin{table*}
\caption{
\label{table:corr:dark_jet}
The correlation coefficients of the logits of the model output, $\mathrm{logit}(\hat{y})$, between the trained models for the dark jet samples.
The coefficients of the same models are the correlation coefficients of the outputs between 
the same networks trained with different random number seeds.
}
\begin{ruledtabular}
\begin{tabular}{cccccccc}
       & MF    & RN    & RN+MF  & CNN   & CNN+MF &  EFN & EFN+ MF \\
\hline 
MF     & 0.976 & 0.681 & 0.801  & 0.736 & 0.780 & 0.609  & 0.712\\
RN     &       & 0.942 & 0.868  & 0.745 & 0.732  & 0.705 & 0.723\\ 
RN+MF  &       &       & 0.973  & 0.793 & 0.839  & 0.679 & 0.777\\
CNN    &       &       &        & 0.958 & 0.924  & 0.763 & 0.809\\
CNN+MF &       &       &        &       & 0.967  & 0.727 &  0.822    \\
EFN    &       &       &        &       &        & 0.902 & 0.873 \\
EFN+MF &       &       &        &       &        &       & 0.933 \\
\end{tabular}
\end{ruledtabular}\label{tableiV}
\end{table*}

As a working example of our network, a toy Hidden Valley model \cite{Strassler:2006im,Carloni:2011kk} whose signature a semi-visible jet \cite{Cohen:2015toa,Bernreuther:2020vhm} is considered.
The hidden sector may include a fermion $q_v$ charged under the secluded gauge group and a massive leptophobic gauge boson $Z'$ that mediates the interaction between the SM particles and the hidden sector.
At the hadron collider, $q_v$ may be produced through the process $q\bar{q}\rightarrow Z' \rightarrow q_v \bar{q}_v$.
The secluded gauge interaction confines $q_v$ and $\bar{q}_v$ and forms pions $\pi_v$ and rho mesons $\rho_v$ after the hidden sector parton shower and hadronization. 
We consider a scenario that only $\rho_v$ leaves visible signatures via the decay $\rho_v \rightarrow q\bar{q}$ while the other mesons are not visible at the detectors.
The resulting semi-visible jet, which we call a dark jet, contains many color-singlet quark pairs fragmenting into hadrons and missing particles.
Therefore, the dark jets have different geometric and hard substructures compared to the QCD jets.

For the simulation of the dark jet, we use \texttt{Pythia 8} \cite{Sjostrand:2014zea} and its Hidden Valley model implementation \cite{Carloni:2011kk}.
The mass spectrum is assigned as follows: $m_{Z'}=1400 \, \mathrm{GeV}$, $m_{q_v} = 10\,\mathrm{GeV}$, and $m_{\pi_v} = m_{\rho_v} = 20 \,\mathrm{GeV}$.
The fraction of $\pi_v$ and $\rho_v$ during the hadronization is 1:3, as the spin counting suggests.
The QCD jet samples are the leading $p_T$ jets of the process $pp\rightarrow 2j$,  and they are generated using \texttt{MadGraph5 2.6.6} \cite{Alwall:2014hca} together with \texttt{Pythia 8}.
Detector effect is modeled by \texttt{Delphes 3.4.1} \cite{deFavereau:2013fsa} with the default ATLAS detector card.

The training and test samples are the leading $p_T$ jets with $p_{T,\jet} \in [150,300]$~GeV and $m_{\jet} \in [30,70]$~GeV.
The number of selected events is $6.0\times 10^5$ for the dark jet samples and $1.9\times10^6$ for the QCD jet samples.

\Figref{fig:ms:a} shows the $A^{(k)}$ distributions of dark jets and QCD jets.
The most left curve is the $A^{(0)}$ distributions, and they are close to each other. 
On the other hand, the average of $A^{(i)}$ ($i>0$) of the QCD jets is much larger, and the $A^{(i)}$ distribution extends far beyond the endpoint of the dark jet $A^{(i)}$ distribution. 
The RN+MF model can {explicitly} use the feature in the classification. 

Given the apparent difference of $A^{(i)}$ distributions, the CNN is also capable of learning this phase space where only QCD jets exist.  
The classifier reasoning appears in the dijet distribution in \figref{fig:ms:b}.
The distributions are after applying the mild cut of 90\%  signal dark jet efficiencies using the CNN. 
The cut significantly suppresses the events beyond the endpoint of the dark jet distribution.

We show the receiver operator characteristic (ROC) curves of RN,\footnote{EFN results are explained in \sectionref{sec:benchmark_tagging:efn}.} and CNN with and without the MFs on \figref{fig:roc:darkjet}.
The corresponding area under the ROC curve (AUC) in \tableref{table:auc:dark_jet}. 
Both RN and CNN models reject more than 90\% QCD jets on the phase space of large MFs without losing any dark jet events as illustrated in \figref{fig:ms}.
Even a simple classifier using only the MFs and kinematical variables rejects most of the QCD jet samples, as seen by the orange curve. 
This shows that the MFs describe the boundary of the phase space of the dark jet events quite efficiently. 
The model with MF consistently outperforms the one without MF, as can be seen in \tableref{table:auc:dark_jet}. 
The AUC of RN+MF is slightly better than CNN, and the AUC of CNN+MF is the best among the CNN and RN models.

The ROC curves show some crossovers in the region of small dark jet tagging efficiency below $\epsilon_{dark}=0.6$, and RN+MF rejection efficiency looks better than CNN+MF in such regions.   
However, the rejection rate is so high that a relatively small training sample of O(1000) events is available for the training. A slight difference in the rejection efficiency is therefore not statistically significant.

We can estimate the difference between the CNN and RN+MF models by calculating the correlation coefficient of the logit outputs $\mathrm{logit}(\hat{y}_{\mathrm{CNN}})$ {and} $\mathrm{logit}(\hat{y}_{\mathrm{RN+MF}})$ for the same testing event set. 
We list the values in \tableref{table:corr:dark_jet}. 
Here $\hat{y}$ is the  outputs of each model, and its logit is  $\mathrm{logit}(\hat{y})=\log(\hat{y})-\log(1-\hat{y})$.  
The correlation coefficient $\rho$ between CNN and RN+MF is relatively small, {and} $\rho=0.793$ for {the dark jet dataset.}
But once we give the MF information to the CNN model, the correlation improves, {and} $\rho=0.893$ between CNN+MF and RN+MF.  
The improvement of correlation and classification performance indicates that the CNN is not fully utilizing those MFs unless explicitly provided as inputs.

The correlation coefficient between the network outputs trained with different random number seeds is significantly larger than the correlation between the different models.
This indicates that the difference between the network outputs is primarily due to the systematic difference in the network architectures.

\subsection{Top Jet Tagging}

\begin{figure}
\includegraphics[width=0.45\textwidth]{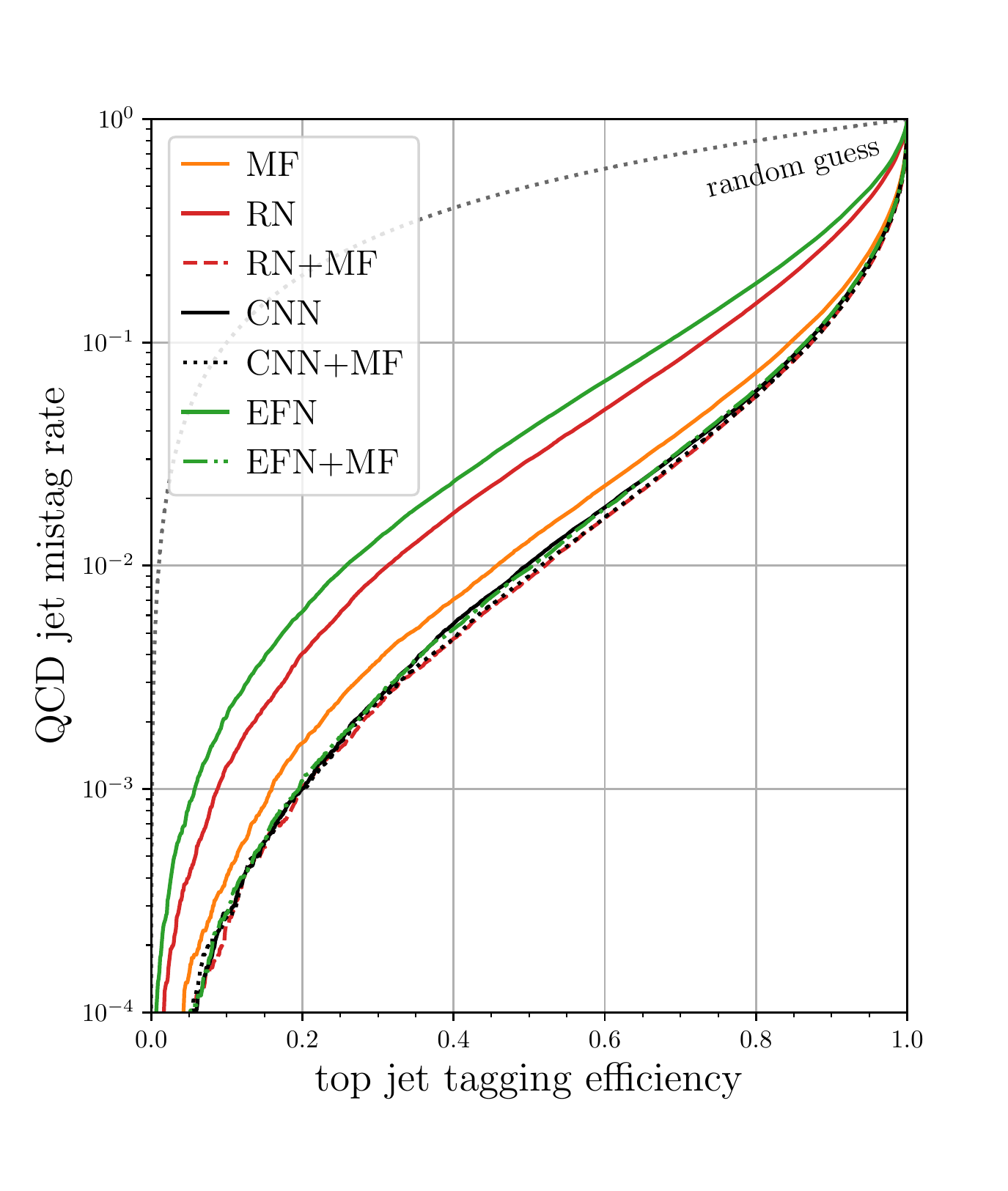} 
\caption{\label{fig:roc:top}
ROC curves of various classification models for top jets vs.~QCD jets.
}
\end{figure}

\begin{table}
\caption{
AUC of various top jet taggers. 
The EFN models have 50 hidden features at the first dense layer.
We also show the training time $t_{\mathrm{train}}$ and the number of epochs at the end of the training, $N_{\mathrm{train}}$ {for} mini-batch numbers $N_{\mathrm{batch}}=$ 20 and 200.
}
\begin{ruledtabular}
\begin{tabular}{lccc}
      & 
      \multirow{2}{*}{AUC}
      &
      \multicolumn{2}{c}{$t_{\mathrm{train}} / N_{\mathrm{epoch}}$}
\\
& & $N_{\mathrm{batch}}=20$ & $N_{\mathrm{batch}}=200$ \\
\hline 
MF      & 0.9467   & 793 s / 564 epochs & \phantom{00}954 s / \phantom{0}363 epochs \\
\hline
RN      & 0.9038   & 288 s / 186 epochs & \phantom{00}619 s / \phantom{0}214 epochs \\
RN+MF   & 0.9552   & 418 s / 255 epochs & \phantom{0}1057 s / \phantom{0}288 epochs \\
CNN     & 0.9529   & & 31020 s / 1483 epochs \\
CNN+MF  & 0.9547   & & 12319 s / \phantom{0}530 epochs \\
\hline
EFN     & 0.8900   & 535 s / 120 epochs & \phantom{00}723 s / \phantom{0}108 epochs \\
EFN+MF  & 0.9521   & 725 s / 149 epochs & \phantom{00}813 s / \phantom{0}111 epochs \\
\end{tabular}
\end{ruledtabular}\label{tablev}
\end{table}

\begin{table*}
\caption{
The correlation coefficients of the logit of outputs between the trained models for top jet samples.
Diagonal elements are the correlation coefficents between the same networks trained with different random number seeds.
}
\begin{ruledtabular}
\begin{tabular}{lccccccc}
       & MF    & RN    & RN+MF  & CNN   & CNN+MF & EFN & EFN+MF\\
\hline 
MF     & 0.990 & 0.670 & 0.922  & 0.808 & 0.924 & 0.635 & 0.911 \\
RN     &       & 0.978 & 0.778  & 0.738 & 0.730 & 0.847 & 0.714\\ 
RN+MF  &       &       & 0.986  & 0.847 & 0.941 & 0.711 & 0.931\\
CNN    &       &       &        & 0.933 & 0.866 & 0.739 & 0.849\\
CNN+MF &       &       &        &       & 0.979 & 0.723 & 0.945\\
EFN    &       &       &        &       &       & 0.913 & 0.727\\
EFN+MF &       &       &        &       &       &    & 0.960 \\
\end{tabular}
\end{ruledtabular}\label{tablevi}
\end{table*}

For the top jet study, we use the samples described in \cite{Chakraborty:2020yfc}.
We use the events with $p_{T,\jet} \in [500,600]$~GeV and $m_{\jet} \in [150,200]$~GeV.
The number of selected events is $9.5\times 10^5$ for top jets and $3.5\times 10^5$ for QCD jets. 
The ratio between training, validation, and test samples and {the training method} is the same as the dark jet case. 

We show the ROC curves in \figref{fig:roc:top}.
The model MF, which uses only the MFs as inputs (without any IRC safe correlators), performs better than the RN model. 
This indicates that the geometric and topological information is the primary information for the top jet classification.
As can be seen in \tableref{tablev}, the model using IRC safe variable with MFs is better than the one without MFs as the dark jet case. 
The MFs are enhancing the performance of the RN much more than the dark jet tagging case.

The CNN+MF shows a similar tagging performance to the RN+MF, but the baseline CNN does not.
As discussed earlier, the convolutional representation of the MFs involves a discontinuous step function.
However, the step function is hard to be modeled by convolutional layers with a finite number of filters and L2 regularizers.
This CNN setup effectively penalizes functions with discontinuity because it requires large weights or a large number of filters with small weights.

The correlation coefficient $\rho$ of the logit of outputs among the training of the same model with different random number seeds is  0.986 for RN+MF.  
On the other hand, the $\rho$ of CNN is 0.933. 
The difference shows that the training of the CNN model suffers the local minimum problem relative to RN+MF.
In gradient-based training methods, easily classifiable samples dominate the early phase of the training.
The different training may show us different local minima that mainly describe the classification boundary for the dominant samples.
In such cases, confusing events are underrepresented, and the training results will have some variance.
This variance is larger for the more generic function model, and the CNNs have a larger correlation coefficient than the RN+MFs.

The local minimum problem of the CNN can be relaxed by explicitly providing some components, such as the MFs.
Adding the MFs to CNN inputs improves {the situation, and CNN+MF has the correlation coefficient 0.979.}
Furthermore, the correlation between CNN+MF and RN+MF is 0.941, much higher than {the correlation between CNN and RN+MF.}
Namely, the two models are now quite {correlated} to each other.

To visualize the fine difference between the RN+MF and CNN, we compare the $(A^{(0)},A^{(2)})$ distribution of dijet samples, conditioned on the classifier outputs.
We select the dijet samples with classifier outputs $\hat{y}_{\mathrm{CNN}}$ and $\hat{y}_{\mathrm{RN+MF}}$ of CNN and RN+MF models less than its value at the 70\% of top jet selection efficiency, respectively.

\begin{figure}
    \hspace*{-1.2em}
    \subfloat[\label{fig:correlation:a}]{
        \includegraphics[height=0.225\textwidth]{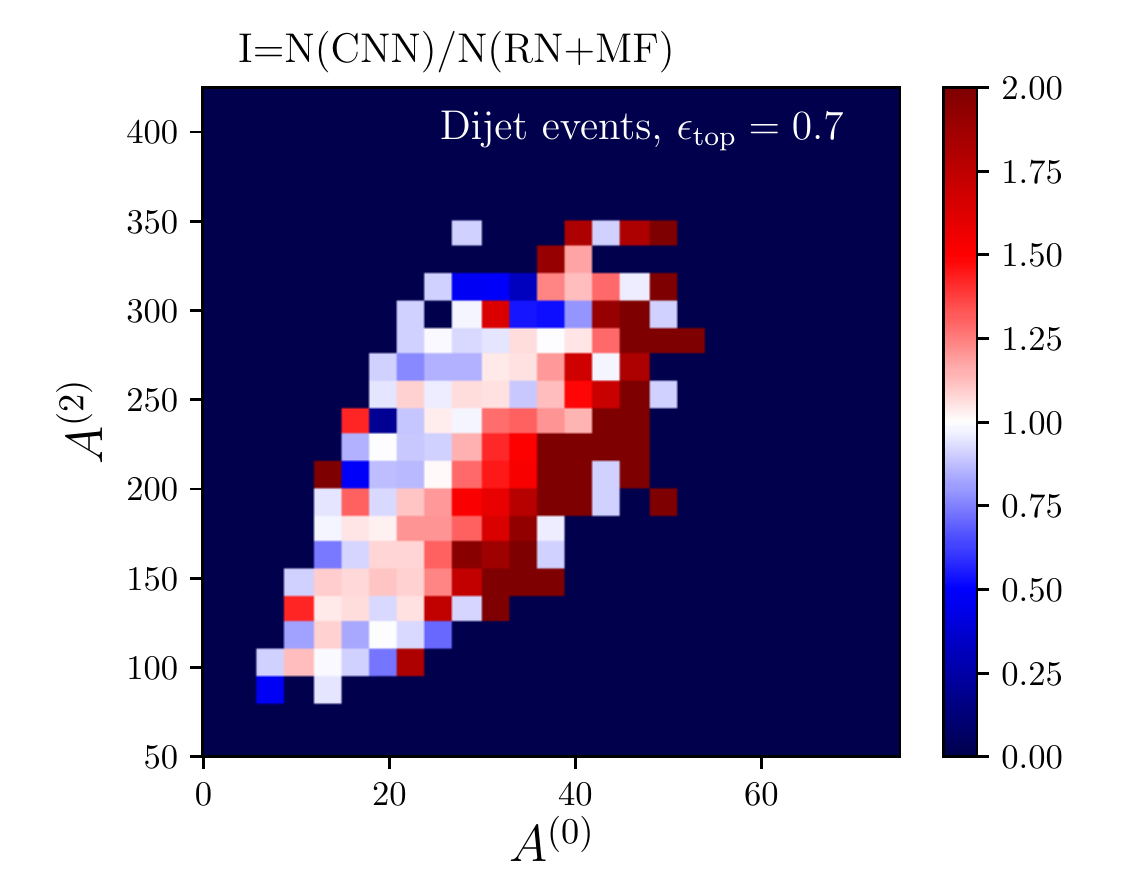}
    }\hspace*{-4.0em}
    \subfloat[\label{fig:correlation:b}]{
        \includegraphics[height=0.225\textwidth]{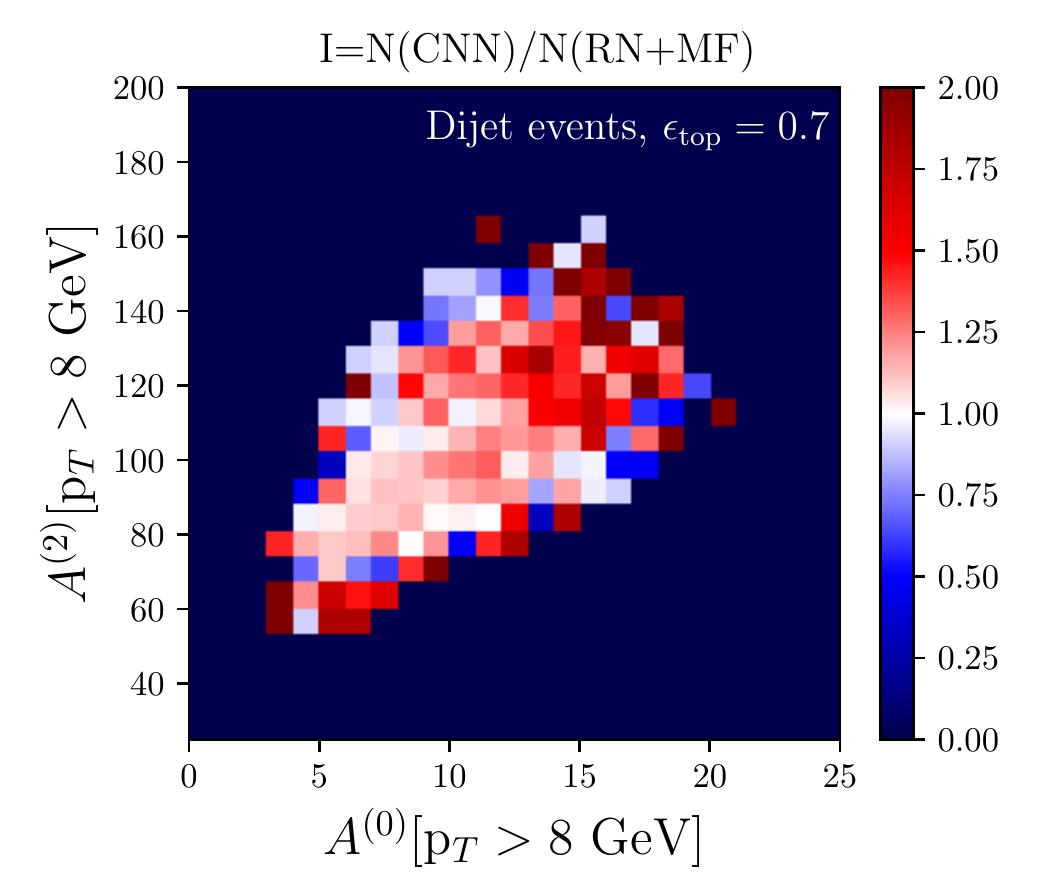}
    }
\caption{\label{fig:correlation} 
The PDF of the dijet event in CNN model  
divided by the one in RN+MF model with the same signal efficiency
at $\epsilon_{top}=0.7$ .  }
\end{figure}

By taking the ratio of the histograms of the MFs, we can visualize the difference in classification boundaries of RN+MF and CNN.
In \figref{fig:correlation}, we consider the ratio
\begin{equation}
\mathcal{I}=\frac{N(\mathrm{CNN})}{N(\textrm{RN+MF})+\epsilon}
\end{equation}
where $N$ is the density at a given bin of the histogram of the samples selected by the CNN or RN+MF, and $\epsilon=0.1$ is the regularization to avoid dividing by zero. 
\Figref{fig:correlation:a} is distribution of $\mathcal{I}$ in  $( A^{(0)},  A^{(2)})$ plane,
and \figref{fig:correlation:a} is the same plot but for the MFs obtained from the pixels above the 8~GeV threshold,  $( A^{(0)}[8~\mathrm{GeV}],  A^{(2)}[8~\mathrm{GeV}])$.

Because the RN+MF model rejects more dijet events, the ratios tend to be bigger than 1 for most of the bins. 
In the figure, the red bins represent  ${\cal I}>1$, while the blue bins correspond to ${\cal I}<1$.  
For \figref{fig:correlation:a}, the bins with large $A^{(0)}$ and small $A^{(2)}$ is red, indicating the RN+MF improves the classification by selecting more samples on this region.
For \figref{fig:correlation:b}, the region with large $A^{(0)}$ and large $A^{(2)}$  tend to have larger values, but the red region is less prominent.
This may indicate that the CNN is utilizing the geometric features of the pixels with energy above 8~GeV, but the CNN may also have difficulty in fully utilizing the geometric information of soft energy deposits.

\subsection{Comment on EFN and EFN+MF }
\label{sec:benchmark_tagging:efn}

In addition to CNN, we study the classification using EFN and EFN+MF models.
The EFN model uses the same jet images as inputs, but the model itself is constrained to be IRC safe.
Because of the constraint, the EFN cannot fully use the geometric information of the soft activities encoded in the MFs.
As a result, the classification performance of EFN is worse than that of the networks taking MFs as inputs and the CNN, which implicitly cover the MFs.
Nevertheless, the EFN+MF works nearly as equal as the CNN+MF and RN+MF, and it covers sufficiently useful IRC safe information for both dark jet tagging and top jet tagging.

In the dark jet tagging, the IRC safe variables are the key information for the jet tagging, and EFN performs well in the classification as illustrated in \figref{fig:roc:darkjet}.
In addition, considering MFs as extra inputs improves the performance slightly.
At the low signal efficiency, the EFN+MF model has the best among all models in \figref{fig:ms}.
As discussed already, due to the large background rejection in the region, the number of the training sample is enough, and {we suspect that} the difference is within the statistical fluctuations.

In the top tagging, the geometric and topological information is important. 
The performance of sole EFN is comparable to that of the RN, but it is significantly improved when MFs are considered as additional inputs.
Our RN model uses the two-point correlation to the leading $p_T$ subjet and two-point correlation after removing the leading subjet to capture the three-point correlation inside the top jet.
The inputs for the EFN are also sensitive to this topological three-prong structure of the top jet because we preprocess the jet images, and those three subjets always appear at particular points on the jet image.
The EFN+MF covers more geometric information than the EFN, and its performance is comparable to the CNN as a result.
But the improved performance mostly comes from the MFs, and the EFN+MF works nearly as equal as the CNN+MF and RN+MF.

\section{Computational Advantages of Morphological Analysis and Relation Network}
\label{sec:training_performance}

\subsection{Overcoming a Small Dataset}
As discussed in the previous section, the RN+MF model has some merit over the CNN model on better training performance. 
Models with broader coverage, such as CNN, are capable of modeling generic functions.
The price of the high expressive power is often the high variance in the trained outputs and the high sensitivity to the statistical noise.
These errors may degrade the generalization performance of the network.
In this respect, using a simpler model helps to maintain the performance for some cases.

\begin{figure}
\includegraphics[width=0.234\textwidth]{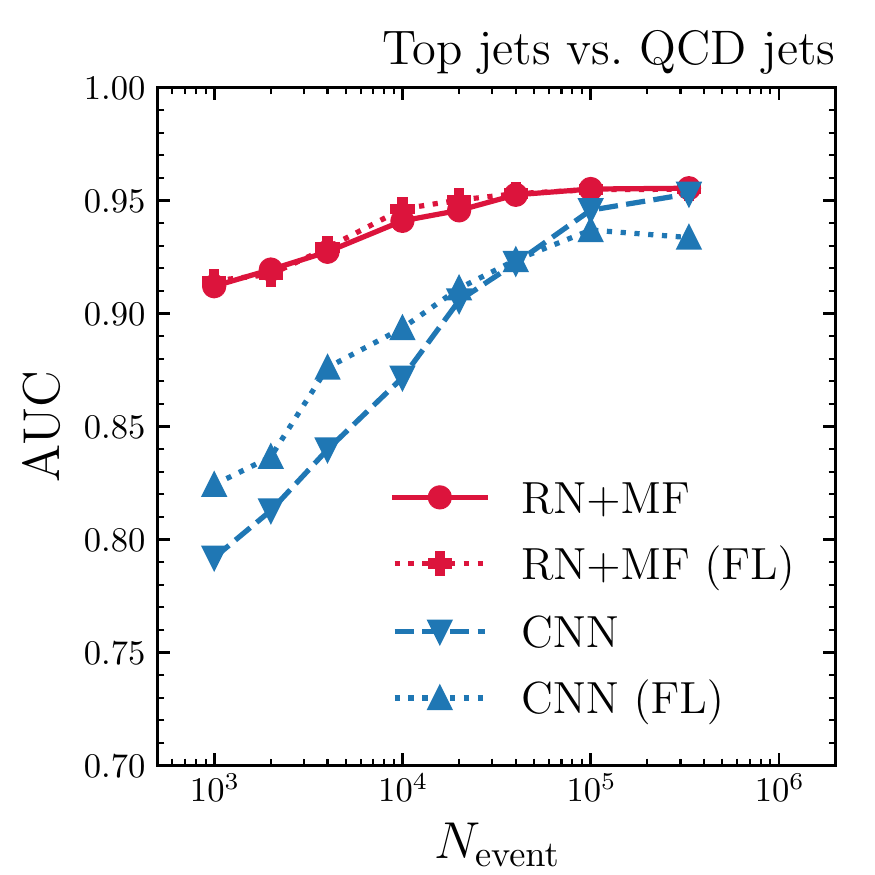} 
\includegraphics[width=0.234\textwidth]{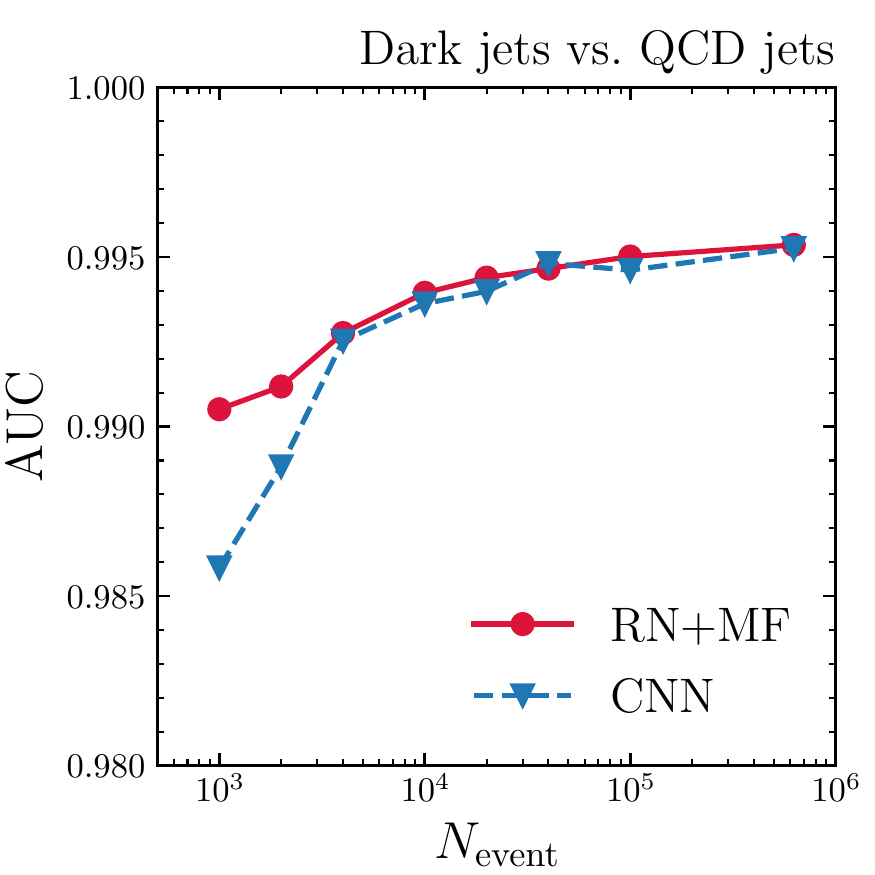}
\caption{\label{fig:auc_nevent}  
The AUCs of RN+MF and CNN trained with a given number of training samples.
The $x$-axis $N_{\mathrm{event}}$ denotes the number of samples in each class.
The rightmost entries are the AUCs of the networks trained on the full training dataset. 
Since the number of the signal and background samples are not identical in this case, we put their average value on the $x$-axis.
}
\end{figure}

\Figref{fig:auc_nevent} shows the AUCs of RN+MF and CNN as the functions of the number of training samples. 
We can see from the figure that the AUC of RN+MF is significantly larger than that of CNN for a small dataset, although their gap decreases as the size of the training dataset increases.

For the top jet classification, RN+MF achieves the AUC higher than 0.9 already at 1000 training samples, and the AUC is only 4\% smaller at most than the best AUC.
Meanwhile, CNN needs $\mathcal{O}[10,000]$ samples to achieve the same performance as RN+MF.

We find similar behavior of the AUC curves in the dark jet classification.
The curves for RN+MF and CNN meet at 4000 events, which is much smaller than the meeting point of the curves in the top jet tagging case.  
This is because there are no dark jet samples at the tail of the MF distributions of QCD jets, as shown in \figref{fig:ms}.
The training of the CNN could easily find this difference with a small number of samples, and the curves will meet much earlier.

Since the CNN model has comparable performance to RN+MF, we may consider optimizing learning steps to improve the performance when the dataset is small. 
For example, we may adjust learning dynamics by replacing the cross-entropy loss $\mathcal{L}_{\mathrm{CE}}$ with a focal loss $\mathcal{L}_{\mathrm{FL}}$ \cite{Lin_2017_ICCV},
\begin{eqnarray}
    \nonumber
    \mathcal{L}_{\mathrm{FL}} 
    & = &
    -\frac{1}{2} \E \left( (1-\hat{y})^2 \log \hat{y} \,|\, y=1 \right)
    \\ \label{eqn:loss_fl}
    & &
    - 
    \frac{1}{2} \E \left( (\hat{y})^2 \log (1-\hat{y}) \,|\, y=0 \right).
\end{eqnarray}
The results are shown in dotted lines in \figref{fig:auc_nevent}. 
The focal loss penalizes the contribution from easily-classifiable examples by extra factors $(1-\hat{y})^2$ and $(\hat{y})^2$, and it helps training when the dataset is sparse.
The jet image dataset is sparse, so that we can see the improvement in the low statistics.
However, there are no improvements to RN+MF since $\MF$ and $S_2$ distributions are mostly dense and smooth.
Note that the training using focal loss does not converge to the maximum likelihood estimatiion of the binary classifier, i.e., $\hat{y} \nrightarrow p(y=1|x)$ in the asymptotic limit.
Therefore, the performance is generally less than the one using the cross-entropy loss when enough data is available.

\subsection{Less Computational Complexity and Training Time}

Another advantage of the RN+MF is its low computation complexity. 
Networks with less computational complexity can be evaluated much faster and takes less memory.

\Tableref{tableiii} and \tableref{tablev} show that the training time of RN+MF is about ten times shorter than that of CNN.
We also note that RN+MF takes about 300 MB GPU memory during the training with 200 mini-batches, while CNN takes about 6000 MB GPU memory in our setup.

We can estimate the computational complexity difference between CNN and RN+MF from the complexities of network evaluations and the input calculations.
Because input calculations can be cached, the network evaluation complexity is the dominant factor to the complexity during the training.
The evaluation complexity is proportional to the number of multiplications since the networks mostly consist of tensor multiplications.
One of the most expensive layers of our CNN is a convolution layer with $3\times3$ filters mapping images with $30\times30$ pixels and 16 channels to the images of the same size. 
This layer has the following number of multiplications,
\begin{equation}
    (3 \times 3) \times (16 \times 16) \times (30 \times 30) = 2,073,600.
\end{equation}
Our CNN has two convolutional layers with this configuration, so that those two layers used about $4,000,000$ multiplications.

Meanwhile, our RN+MF has only fully connected layers, and the most expensive one has 200 incoming features and 200 outgoing features. 
This layer has $200 \times 200 = 40,000$ multiplications.
We use three dense layers for each of the MLPs of RN+MF, which have four MLPs. 
Then the number of multiplications is at most
\begin{equation}
    3\times4\times40,000 = 480,000.
\end{equation}
The estimated computational complexity is factor 10 less than the convolutional layers, and it qualitatively explains the difference in training time.
It also explains the difference in GPU memory usage since the backpropagation algorithm has to record the entire operations.
More operation is involved, more GPU memory is needed during the training.

On the other hand, the complexity of input calculations only matters when the network inputs are not cached. 
The computational complexity of evaluating the inputs of RN+MF is as following.
The calculation of MFs has two convolutions with filter sizes $(2k+1) \times (2k+1)$ and $2 \times 2$ for the dilation and local feature identification, respectively. 
Those two convolutions have the number of multiplications,
\begin{equation}
    (2k+1) \times (2k+1) \times (30 \times 30) +  (2\times 2) \times (30 \times 30),
\end{equation}
which is 4,500 for $k=0$ and 155,700 for $k=6$.
Note that the complexity of dilation, $(2k+1) \times (2k+1) \times (30 \times 30)$, can be further reduced by using optimized algorithms.
We may consider this number as the upper bound of the complexity.

The calculation complexity of the two-point correlation $S_{2,ab}$ is a function of the number of jet constituents, $N$.
The jet reclustering has $N \log N$ complexity \cite{Cacciari:2005hq}, and the two-point correlation calculation has $N^2$ complexity in general.
In the case of $N=50$, which is approximately the largest number of jet constituents in our sample according to \figref{fig:ms}, the total complexity is $ \approx 2,700$.
The second $N^2$ factor can be reduced to $N^2 / 2$ if $a$ and $b$ of $S_{2,ab}$ are the same.

Those two complexities of evaluating the inputs of RN+MF, 155,700 and 2,700, are still much smaller than the complexity of the two convolutions layers.
We conclude that the RN+MF setup is computationally efficient than the CNN.

\section{Parton Shower Modeling and Minkowski Functionals}
\label{sec:ps_and_mf}

So far, we have been discussing jets generated by \pythia{}, but the predicted jet substructure has a simulator dependency in general because of different parton shower schemes.
\pythia{} adopts $p_T$-ordered showering \cite{Rasmussen:2015qgr,Corke:2010yf} while \herwig{} adopts angular-ordered showering.
The distributions of MFs with energy thresholds can capture the geometric differences between those two shower schemes, and the two simulated distributions may be different from each other.
We quickly check the difference in $A^{(k)}[p_T]$ distributions and discuss the origin of difference in terms of the shower scheme.

\begin{figure}[tb]
        \hspace*{-1.0em}
        \subfloat[\label{fig:pyhw:a}]{
            \includegraphics[width=0.27\textwidth]{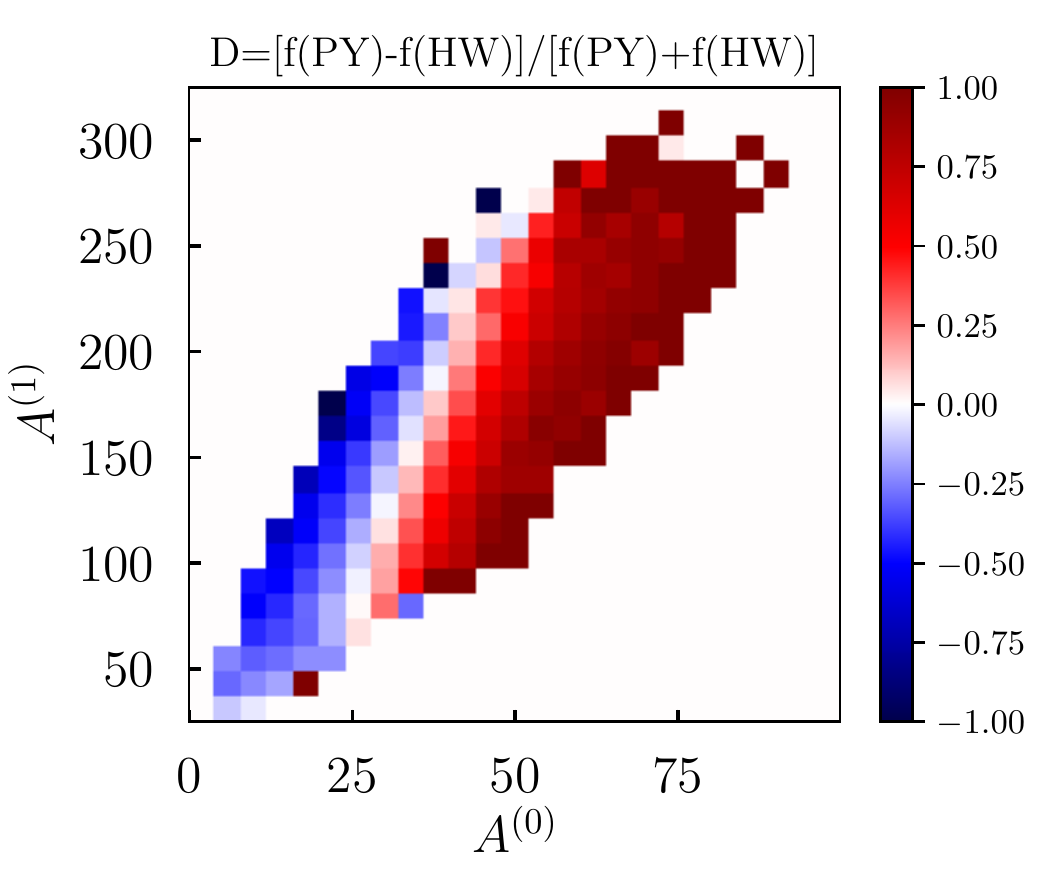}
        }
        \hspace*{-3.4em}
        \subfloat[\label{fig:pyhw:b}]{
            \includegraphics[width=0.27\textwidth]{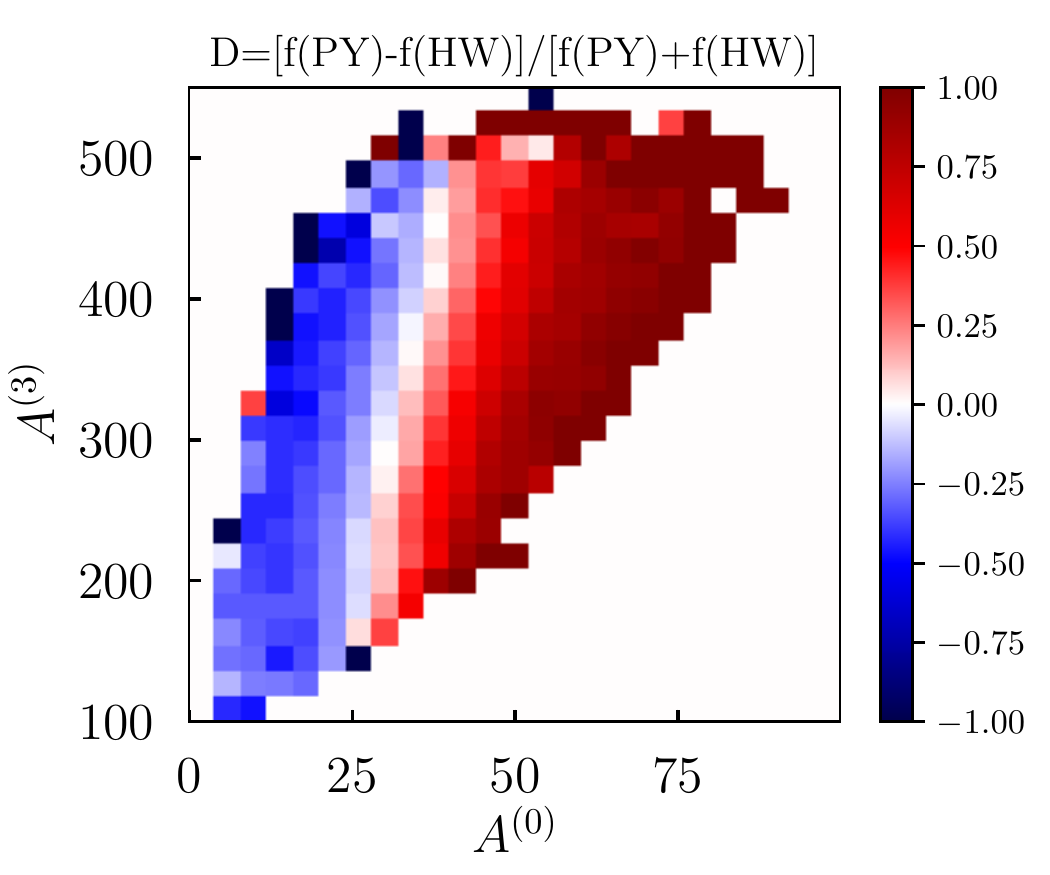}
        }
        
        \hspace*{-1.0em}
        \subfloat[\label{fig:pyhw:c}]{
            \includegraphics[width=0.27\textwidth]{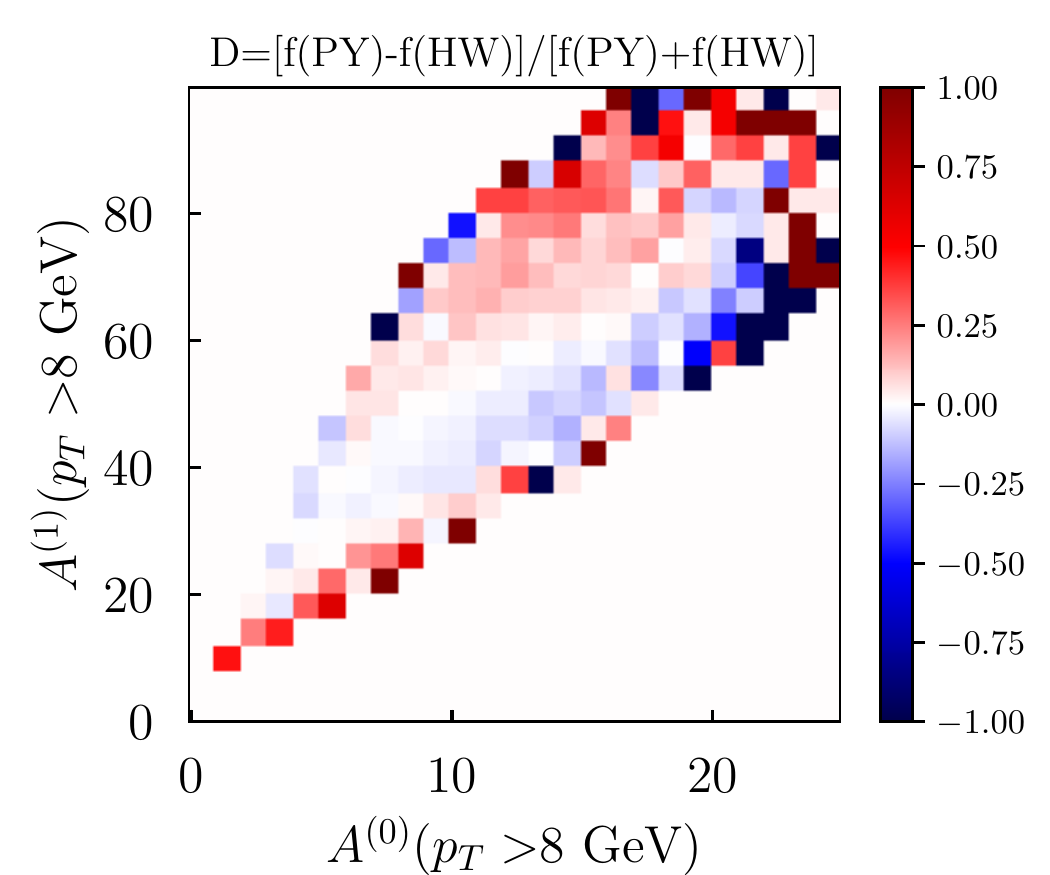}
        }
        \hspace*{-3.4em}
        \subfloat[\label{fig:pyhw:d}]{
            \includegraphics[width=0.27\textwidth]{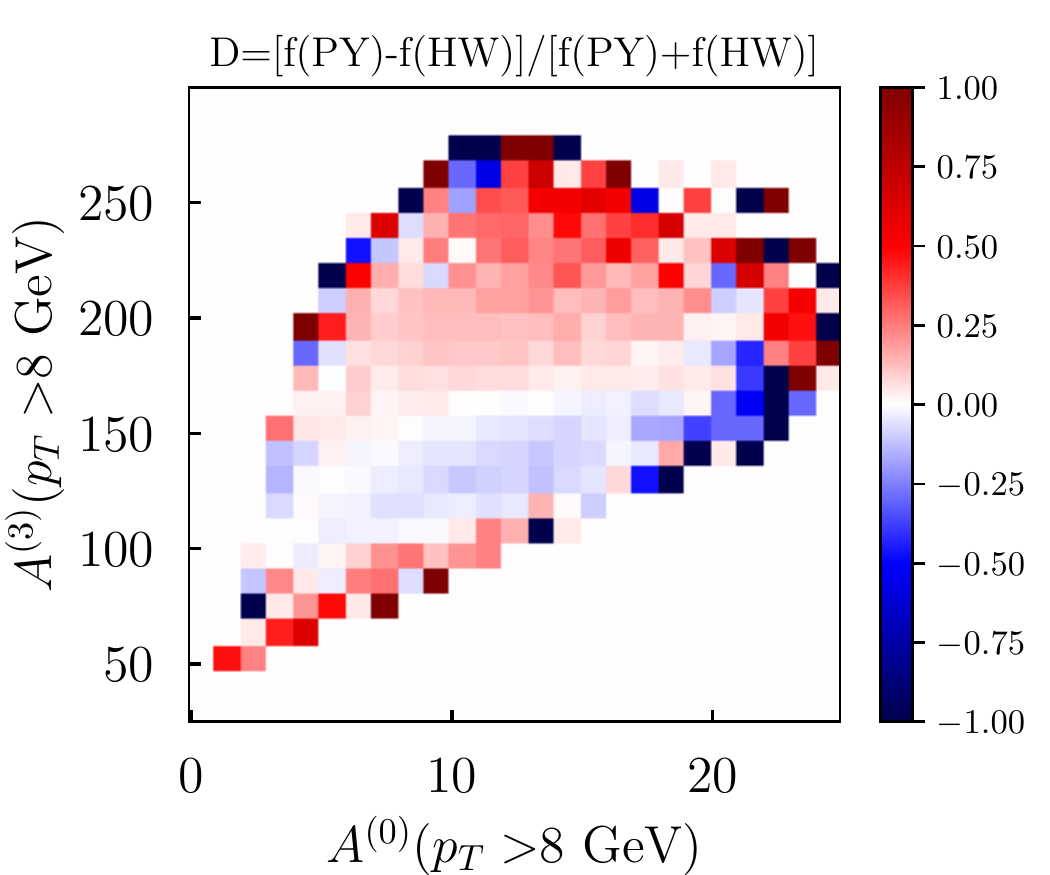}
        }
    \caption{\label{fig:pyhw}  
    The asymmetry $\mathcal{D}$ of the $(A^{(0)}, A^{(k)})$ distributions simulated by \pythia{} and \herwig{}.
    Figures (a) and (c) show the asymmetry of $(A^{(0)},A^{(1)})$ distributions.  
    Figures (b) and (d) show the asymmetry of $(A^{(0)}, A^{(3)})$ distributions. 
    No $p_T$ filter is applied to (a) and (b), while $p_T> 8$~GeV filter is applied for (c) and (d). 
    }
\end{figure}

In \figref{fig:pyhw}, we show the following asymmetry ratio $\mathcal{D}$ of the distribution of two selected $A^{(k)}[p_T]$.
\begin{equation}
    \mathcal{D}(i) = \frac{f_P(i) - f_H(i)}{f_P(i)+ f_H(i)}, \quad f_A(i) = \frac{N_A(i)}{\sum_i N_A(i)} \; \mathrm{for} \;A\in\{P,H\}
\end{equation}
where $N_P(i)$ and  $N_H(i)$ are the number of \pythia{} and \herwig{} events in the $i$-th bin, and $f_P(i)$ and  $f_H(i)$ are its fraction with respect to the total number of events,  respectively.
Here, the samples are the QCD jets of the top jet classification, with $p_{T,\jet} \in [500,600]$~GeV and $m_{\jet} \in [150,200]$~GeV.

In \figref{fig:pyhw:a} and \figref{fig:pyhw:b}, we show the asymmetry ratio of $(A^{(0)}, A^{(1)})$ without $p_T$ filters.
The darkest red bins has $\mathcal{D}=1$, where no \herwig{} events are observed. 
The darkest blue region corresponds to $\mathcal{D}=-1$, and no \pythia{} samples are in there.
The dark red pixels tend to be in large $A^{(0)}$ region, because \pythia{} predicts higher $A^{(0)}$.
For the same $A^{(0)}$ value, \pythia{} predicts smaller values of $A^{(1)}$ than \herwig{}. 
This means the jet constituents are more clustered in \pythia{}.
The trend is common for all $k>1$ (See \figref{fig:pyhw:b} for  $k=3$.) 

The situation is different for $A^{(k)}$ with $p_T$ filter. 
As illustrated in \figref{fig:pyhw:c} and \figref{fig:pyhw:d},
the $A^{(k)}[8\;\mathrm{GeV}]$ of \pythia{} tend to be higher than that of \herwig{} for given $A^{(0)}[8\;\mathrm{GeV}]$. 
This means high $p_T$ pixels are more sparsely distributed in \pythia{} generated samples.

Recall that \pythia{} adopts a transverse-momentum-ordered evolution scheme.
A high $p_{\perp}$ radiation in \pythia{} tends to be emitted at a larger angle.
For the case of \herwig{}, the first emission in the evolution is typically a large angle soft radiation.
The asymmetry $\mathcal{D}$ for $A^{(k)}[p_T]$ distributions is consistent with the expectation of the shower modeling. 
\herwig{} QCD jet emits soft particles at a large angle while \pythia{} QCD jet emits higher $p_T$ objects at a large angle.

For the best classification performance with less simulator bias in the application stage,
the distribution of inputs, especially the MFs, has to be tuned carefully to the real experimental data.
The calibration of MF distributions will be helpful to reduce the simulator dependency in the prediction of more general models, such as the CNN, because the MFs are important features in the jet classifications, as shown in \sectionref{sec:benchmark_tagging}.

\section{Summary}
In this paper, we introduce a neural network covering the space of ``valuations" of jet constituents.
The valuations introduced in this paper can be considered as a generalization of particle multiplicities which is a useful variable in quark vs. gluon jet tagging, but it is not IRC safe in general. 
The space of IRC unsafe variables is less explored compared to that of IRC safe variables because of its theoretical difficulties.
Nevertheless, Hadwiger's theorem in integral geometry tells us some structure of the valuation space, which is an interest to this paper.
The dimension of the valuation space is finite, and its basis is called the Minkowski functionals (MFs).
In the two dimensional Euclidean space, the MFs are Euler characteristic, perimeter, and area.
We utilized these geometric features to build a neural network covering the space of valuations, and the resulting network is a multilayer perceptron taking the MFs as inputs.

In the case of jet image analysis, we showed that the MFs of dilated jet images could be represented by a chain of convolutional layers.
Therefore, convolutional neural networks (CNN) can explicitly utilize this information.
In the semi-visible jet tagging example, the CNN finds out the phase-space region of MFs where only QCD background can be found without difficulties.
However, the MFs is not a smooth function of jet images, and the CNN had a problem accessing that information when $L_2$ regularization is involved.
By explicitly adding the MFs as inputs to the CNN, we showed that its classification performance is improved.

We further build up a neural network architecture combining these valuations to the IRC safe information.
In particular, we consider energy correlator based networks: the relation network and the energy flow network.
We combine the outputs from those IRC safe neural networks to the network covering IRC unsafe MFs. 
This combined setup has a comparable performance to the CNN.

The combined model is constrained compared to the CNN, but its classification performance is similar; moreover, it has computational advantages.
First, it has a smaller computational complexity than the CNN so that its evaluation is fast and less memory-demanding.
Second, constrained model generally requires a less number of training samples in order to reach its best performance. 
This network is especially useful when data is expensive.

Since MFs can be embedded to the CNN, they could potentially be interpreting variables of the CNN.
Deep neural networks are a highly expressive model of a function, but their prediction is not explainable \cite{xie2020explainable, 10.1007/978-3-030-76657-3_1} in general.
If we are aware of potentially important features for modeling, we may distill the features \cite{hinton2015distilling,xie2020explainable} by using interpretable models built from the important features in order to get an insight.
It will also allow us to control the network predictions systematically by using domain-specific knowledge.
We built a network based on MFs, which have clear geometric interpretations, and this type of network combined with interpretable IRC-safe neural networks \cite{Komiske:2018cqr,Chakraborty:2019imr} can be an answer for that in jet tagging problems.

For example, the distributions of IRC unsafe variables, including the MFs, have to be appropriately tuned in order to reduce the simulation bias.
Tuning the distribution of jet constituents themselves for that purpose is not trivial because parton shower simulations are approximation and they do not fully cover the phase space of radiated particles.
The expression of the valuation space using MFs is significantly small in dimensions 
and includes important counting variables that also also have geometric meanings. 
Tuning the distribution of MFs by reweighting \cite{Chakraborty:2020yfc,Diefenbacher:2020rna} can be a more feasible method for controlling {systematical errors of modeling} the space of IRC unsafe features.

Finally, although we limit our discussion to the pixelated image analysis, but it would also be interesting to develop a continuum version of this morphological analysis in order to compare it with graph convolutional neural networks \cite{Qu:2019gqs}.
We will leave these interesting possibilities in future studies.

\begin{acknowledgements}
The authors thank to Benjamin Nachman, David Shih, Iftah Galon, Kyoungchul Kong, Mengchao Zhang, Myeonghun Park, and Takeshi Tsuboi for useful discussions.
This work is supported by the Grant-in-Aid for Scientific Research on Scientific Research B (No.~16H03991, 17H02878) and Innovative Areas (16H06492);
World Premier International Research Center Initiative (WPI Initiative), MEXT, Japan. The work of SHL was also supported by the US Department of Energy under
grant DE-SC0010008.
\end{acknowledgements}

\appendix

\section{Network Configurations}
In the following, we show the hidden layer configurations of the networks studied in this paper.
The activations of all the layers are ReLU, except the last dense layer, whose activation is linear.

\subsection{Valuation model and Relation Network}
We model the morphological analysis and the relation network (RN) by MLPs taking MFs and $S_{2,ab}$ as inptus, respectively.
The configuration of the MLP is as follows,
\begin{itemize}
    \item Concatenate inputs and $\xkin$.
    \item \texttt{Dense}: output size: 200,
    \item \texttt{Dense}: output size: 200,
    \item \texttt{Dense}: output size: 5,
\end{itemize}
where \texttt{Dense} is a fully connected layer of given output size.
Note that the first dense layer is essentially the model for the valuation or two-point energy correlations.

\subsection{Convolutional Neural Network}
The baseline CNN is modeled as follows.
\begin{itemize}
    \item \texttt{Conv2D}: filter size: $3\times 3$, 16 filters
    \item \texttt{Conv2D}: filter size: $3\times 3$, 16 filters
    \item \texttt{Conv2D}: filter size: $3\times 3$, 16 filters
    \item \texttt{MaxPooling2D}: pool size: $2\times 2$
    \item \texttt{Conv2D}: filter size: $3\times 3$, 8 filters
    \item \texttt{Conv2D}: filter size: $3\times 3$, 8 filters
    \item \texttt{Conv2D}: filter size: $3\times 3$, 8 filters
    \item \texttt{MaxPooling2D}: pool size: $2\times 2$
    \item \texttt{Dense}: output size: 200
    \item \texttt{Dense}: output size: 10
\end{itemize}
where \texttt{Conv2D} is a convolutional layers and \texttt{MaxPooling2D} is a max pooling layer for two-dimensional pixelated images.
Zero padding is used for calculating convolutions at the pixels near the boundary.
We also showed that this configuration has a similar classification performance to the \texttt{ResNet} and \texttt{ResNeXt} within our setup and training samples \cite{Chakraborty:2020yfc}.

\subsection{Energy Flow Network}
The energy flow network presented in this paper is essentially the MLP of jet images.
However, 900 inputs are much larger than that of the MFs and $S_{2,ab}$, we compress the inputs to 50 (or 200) latent dimensions first.
\begin{itemize}
    \item Concatenate inputs and $\xkin$.
    \item \texttt{Dense}: output size: 50 (or 200)
    \item \texttt{Dense}: output size: 200
    \item \texttt{Dense}: output size: 200
    \item \texttt{Dense}: output size: 10
\end{itemize}
Again, the first dense layer is essentially the model for the linear energy correlators.

\subsection{Multilayer Perceptron Classifier and Logistic Regression}
The selected network outputs are then combined to the binary classifier, i.e., MLP followed by logistic regression.
\begin{itemize}
    \item Concatenate all the inputs and $\xkin$.
    \item \texttt{Dense}: output size: 200
    \item \texttt{Dense}: output size: 200
    \item \texttt{Dense}: output size: 1
\end{itemize}
the final output is trained by minimizing the cross-entropy loss in \eqref{eqn:loss_ce} or the focal loss in \eqref{eqn:loss_fl}.

\section{Comment on Smooth Activation Functions}
\begin{figure}
\includegraphics[width=0.45\textwidth]{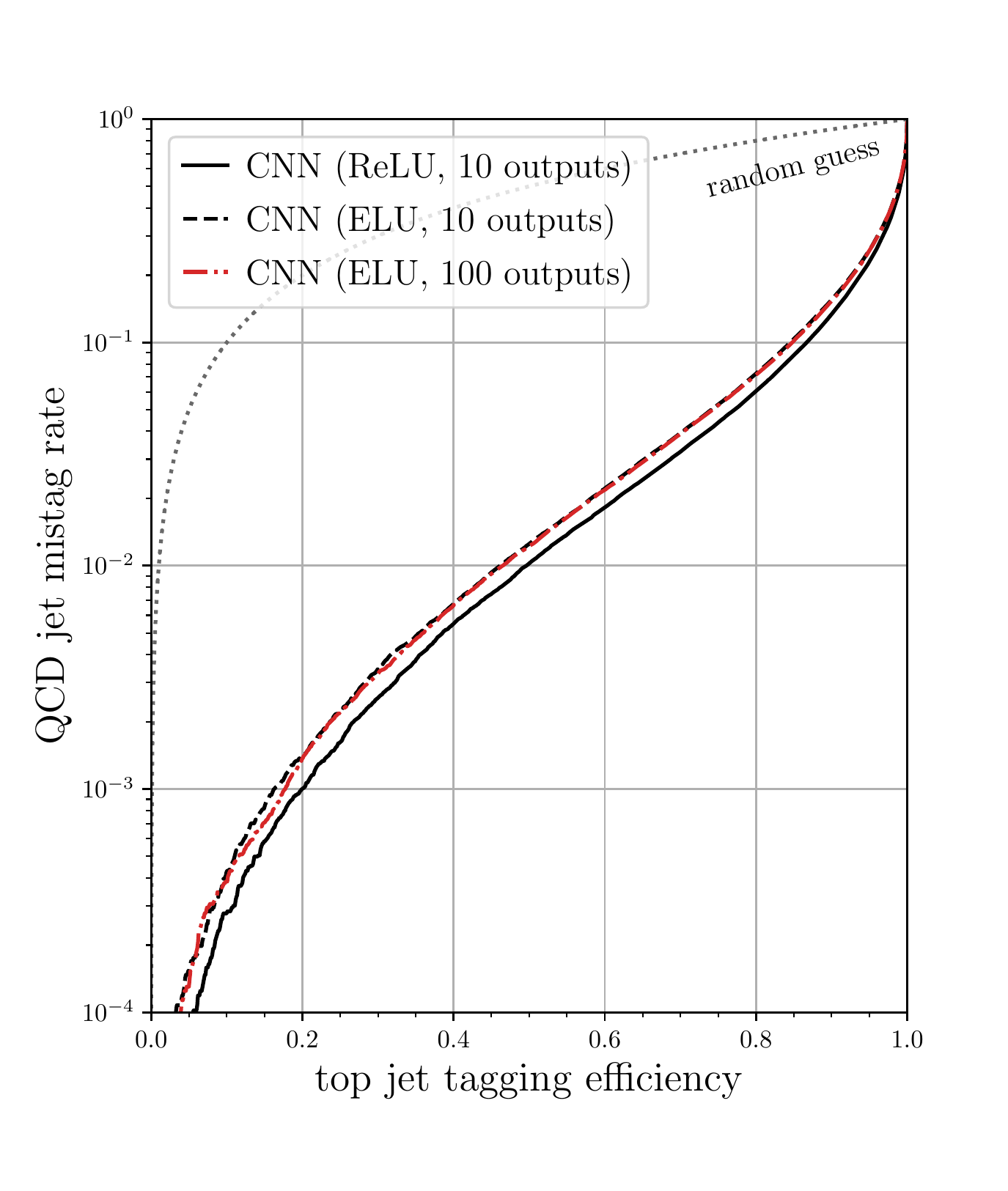} 
\caption{\label{fig:roc:top:elu}
ROC curves of CNNs for top jets vs. QCD jets.
The black solid line is the baseline CNN with ReLU activation in this paper.
Other CNNs use ELU activations.
The red dot-dashed line is the ROC curves of the CNN in \cite{Chakraborty:2020yfc}.
We also show the number of hidden outputs at the last dense layer of the CNN.
}
\end{figure}

In the previous paper \cite{Chakraborty:2020yfc}, we compared the RN with $A^{(0)}$ and $A^{(1)}$ with CNN with ELU activation function and found that the performance is comparable, but this is accidental.
As shown in \figref{fig:roc:top:elu}, the performance of the CNN with ReLU is better than the CNN with ELU \cite{DBLP:journals/corr/ClevertUH15} because ReLU is not a smooth function and can model the step function better.  
Nevertheless, the performance of RN+MF also improves after fully considering the MFs, and the performance is comparable with CNN with ReLU activation, as shown in the main text.

\bibliography{JetSubstructureBib} 

\end{document}